\documentclass[twocolumn,twocolappendix]{openjournal}

\setcitestyle{authoryear,round}
\usepackage[T1]{fontenc}
\usepackage{subcaption}
\usepackage{aecompl}
\usepackage{multirow}
\usepackage{newtxtext,newtxmath}
\usepackage{mdframed,bm}
\usepackage[T1]{fontenc}
\usepackage{ae,aecompl}
\usepackage{graphicx}	
\usepackage{amsmath}	
\usepackage{booktabs}
\usepackage{color}
\usepackage{enumitem}

\usepackage{multirow}
\usepackage[pdftex,dvipsnames]{xcolor}

\usepackage{hyperref}
\hypersetup{
    colorlinks=true,
    citecolor=blue,     
}

\usepackage{physics}
\usepackage{cleveref}
\usepackage{comment}
\usepackage{array}
\usepackage{slashed}
\usepackage[dvipsnames]{xcolor}
\usepackage{xspace}
\usepackage{xcolor}
\usepackage{float}
\restylefloat{table}
\usepackage{physics}
\usepackage[normalem]{ulem}

\usepackage[utf8]{inputenc}
\usepackage[T1]{fontenc}
\usepackage{epsfig,times,amssymb,amsmath,verbatim,xspace}
\usepackage[usenames,dvipsnames,svgnames,table]{xcolor}
\usepackage{tablefootnote}
\usepackage{todonotes}
\usepackage{tabularx}
\usepackage{hyperref}
\usepackage{float}
\usepackage{hhline}
\usepackage{lineno}
\usepackage{multirow}
\usepackage{natbib,ifthen,soul}
\usepackage{orcidlink}
\usepackage{amsmath}
\usepackage{icomma}
\usepackage{academicons}
\newcommand{\orcid}[1]{\href{https://orcid.org/#1}{\textcolor[HTML]{A6CE39}{\aiOrcid}}}

\newcommand{\omegam}{\ensuremath{\Omega_\mathrm{m}}}
\newcommand{\omegab}{\ensuremath{\Omega_\mathrm{b}}}

\newcommand{\as}{\ensuremath{A_\mathrm{s}}}
\newcommand{\ns}{\ensuremath{n_\mathrm{s}}}
\newcommand{\lcdm}{$\Lambda$CDM}

\renewcommand{\arraystretch}{1.35}

\newcommand{\blockfont}[1]{{\textsc{#1}}\xspace}

\newcommand{\argmin}{\arg\!\min}

\DeclareMathOperator{\arcsinh}{asinh}

\usepackage{eso-pic}

\usepackage{orcidlink}
\usepackage{amsmath}
\usepackage{icomma}
\usepackage{academicons}

\shorttitle{Enhancing weak lensing redshift distribution characterization by optimizing the SOMPZ method}
\shortauthors{Campos, et al.\ (DES Collaboration)}

\begin{document}

\AddToShipoutPictureBG*{%
  \AtPageUpperLeft{%
    \hspace*{18cm}%
    \raisebox{-10\baselineskip}{%
      \makebox[0pt][l]{\textnormal{DES-2023-0813}}
 
}}}%

\AddToShipoutPictureBG*{%
  \AtPageUpperLeft{%
    \hspace*{20.25cm} 
    \raisebox{-11\baselineskip}{%
      \makebox[0pt][r]{\textnormal{FERMILAB-PUB-24-0439-PPD}}
}}}%

\title[Enhancing weak lensing redshift distribution characterization by optimizing the SOMPZ method]{Enhancing weak lensing redshift distribution characterization by optimizing the Dark Energy Survey Self-Organizing Map Photo-\textit{z} method\vspace{-0.5cm}}
\author{A.~Campos$^{1,2}$\orcidlink{0000-0002-5124-0771},
B.~Yin$^{3}$,
S.~Dodelson$^{1,2}$,
A.~Amon$^{4,5}$,
A.~Alarcon$^{6,7}$,
C.~S{\'a}nchez$^{8}$,
G.~M.~Bernstein$^{8}$,
G.~Giannini$^{9,10}$,
J.~Myles$^{11}$,
S.~Samuroff$^{12}$
% ...
\vspace{-0.5cm}
\begin{center} DES Collaboration \end{center}
}

\email{andresarodriguescampos@gmail.com}

\begin{abstract}
%
Characterization of the redshift distribution of ensembles of galaxies is pivotal for large scale structure cosmological studies. In this work, we focus on improving the Self-Organizing Map (SOM) methodology for photometric redshift estimation (SOMPZ), specifically in anticipation of the Dark Energy Survey Year 6 (DES Y6) data. This data set, featuring deeper and fainter galaxies than DES Year 3 (DES Y3), demands adapted techniques to ensure accurate recovery of the underlying redshift distribution.
We investigate three strategies for enhancing the existing SOM-based approach used in DES Y3: 
1) Replacing the Y3 SOM algorithm with one tailored for redshift estimation challenges;
2) Incorporating \textit{g}-band flux information to refine redshift estimates (i.e. using \textit{griz} fluxes as opposed to only \textit{riz});
3) Augmenting redshift data for galaxies where available.
These methods are applied to DES Y3 data, and results are compared to the Y3 fiducial ones. Our analysis indicates significant improvements with the first two strategies, notably reducing the overlap between redshift bins. 
By combining strategies 1 and 2, we have successfully managed to reduce redshift bin overlap in DES Y3 by up to 66$\%$. Conversely, the third strategy, involving the addition of redshift data for selected galaxies as an additional feature in the method, yields inferior results and is abandoned. 
Our findings contribute to the advancement of weak lensing redshift characterization and lay the groundwork for better redshift characterization  in DES Year 6 and future stage IV surveys, like the Rubin Observatory.  


\end{abstract}



\section{Introduction}\label{sec:intro}

Large galaxy surveys afford us promising opportunities to learn about the constituents of the universe and the way they are distributed in space. 
This in turn can help us connect fundamental physics -- for example of dark energy and dark matter -- to observations and to learn about the nature of the most mysterious substances postulated to exist. 
Photometric surveys can capture images of many more galaxies than spectroscopic surveys but are hindered by the inability to measure accurate distances to the objects they image. 
{\it Photometric redshifts}, or distances inferred from the observed galaxy properties such as colors, have become essential in extracting information about cosmology from these large surveys.

One of the observables for which photometric redshifts play a major role in is weak gravitational lensing.
Weak gravitational lensing is a fundamental cosmological probe that enables the investigation of the large-scale structure of the universe and has been employed in many contemporary analyses (see, e.g., \citealt*{heymans13,sv-cosmicshear,DLS,hildebrandt17,y1-cosmicshear,hikage19, hamana20,asgari20,loureiro21,DES:2021bvc}; \citealt*{DES:2021vln};  \citealt*{y3-cosmicshearcls}; \citealt*{hsc-1}; \citealt*{hsc-2}). 

In photometric surveys, while galaxy positions serve as tracers of matter density, it is by measuring the distortions in the shapes and orientations of background galaxies induced by the gravitational influence of intervening mass distributions that we can obtain a direct connection to the underlying density field.
However, to extract precise cosmological information from weak lensing, it is imperative to have a robust characterization of the redshift distribution, $n(z)$, of the observed galaxies.
Measuring the spectrum of each galaxy in a large optical imaging survey, though, is unfeasible, and therefore spectra are available only for small subsets of galaxies. 
As a result, photometric surveys heavily rely on limited, noisy photometric bands to estimate redshifts. 
The main challenge arises from degeneracies in the color-redshift relation, which prevent the unique determination of redshifts from wide-band photometry. 
In addition, since lensing samples tend to be fainter/deeper/bluer, they can not be characterised as accurately as the sort of bright red galaxies typically used as lens samples. 
Therefore, the accurate characterization of the redshift distribution thus becomes one of the main challenges, and yet a crucial aspect, for interpreting gravitational lensing measurements, including cosmic shear and galaxy-galaxy lensing correlation functions.

Techniques to estimate photometric redshifts date back several decades.
Template-fitting methods compare the observed photometric data of galaxies with a library of template spectra, allowing for redshift estimation \citep*{Benitez2000, Ilbert2009}. However, this approach can be sensitive to template choices and might not capture all spectral features accurately, leading to biases in redshift predictions, particularly for poorly represented galaxy populations. Empirical approaches exploit statistical correlations between observable features (e.g., color-redshift relations) to estimate photometric redshifts \citep*{BlakeBridle2005, Mandelbaum2008}. However, these methods necessitate accurate and extensive spectroscopic data for calibration. Machine learning techniques, such as artificial neural networks or random forests, have gained popularity due to their ability to learn complex photometry-redshift relationships from training datasets \citep*{CollisterLahav2004, KindBrunner2013}. Nonetheless, these methods heavily rely on the quality and representativeness of the training data, and their performance can degrade when extrapolating to redshift regimes not adequately covered by the training set.
Most recently, unsupervised machine learning methods that compress data embedded in a Bayesian approach have emerged as a promising direction~\citep[see for example,][]{DES:2019bxr}.

A Self-Organizing Map (SOM), also known as a Kohonen map \citep*{SOM:1982}, is an unsupervised machine learning algorithm and neural network architecture
used for dimensionality reduction and data mining. It allows for complex and high-dimensional data to be represented in a lower-dimensional space while preserving the topological relationships between data points. For the purposes of redshift estimation, when assigning each galaxy to a cell in a Self Organizing Map (SOM),
galaxies with similar redshifts are grouped in the same cell, or ``nearby'' cells if the grouping is in a 2D grid, and the redshift distribution for those galaxies can be determined fairly accurately. DES used this technique in its analyses of the data from the first three years of observations, i.e., the Y3 weak lensing cosmological analyses~\citep*{DES:2020ebm,DES:2021wwk,DES:2021bvc,DES:2021vln}, and also as an additional validation follow-up of the Y3 lens sample calibration \citep*{y3-2x2ptaltlenssompz}. 
The KiDS collaboration has also used it \citep*{Wright_2020} to achieve few-percent level constaints on the mean of the redshift distribution for each redshift bin.
It has emerged as a viable candidate for upcoming surveys such as Rubin and Euclid  \citep*{lsst,euclid}, but improvements are required to achieve sub-percent level constraints \citep*{lsstRequirements,euclidRequirements}. 

Here, we explore several improvements to the SOM methodology used in DES-Y3, ahead of the final DES Year-6 (Y6) analysis. This serves two primary purposes: (i) allowing for the potential of improving on the Y3 implementation and (ii) stress-testing the robustness of the cosmological conclusions. The latter point is particularly important in the context of more stringent requirements that come with more statistically powerful data, as well as applying this methodology to deeper photometric data. If different implementations of the SOM framework give the same answer, we will become more confident applying it moving forward as statistical errors continue to decrease.


First, we test replacing the SOM algorithm used in Y3 by the one proposed in \citet*{Sanchez2020}. This new algorithm implements a Self-Organizing Map with a distance metric specific for the problem of photometric redshift estimation. Although it was shown in the Year 3 analysis that the generic SOM algorithm is already successful at estimating redshifts at the percentage level \citep*{DES:2020ebm}, we hope that by introducing a SOM that is tailored for the problem of redshifts, we can achieve even better precision. Second, we show the impact that including an extra flux band, the \textit{g}-band, has on our ability to obtain well-defined redshift bins, motivating the importance of well calibrated point spread functions in those limits, such that we do not lose this very crucial piece of information. Finally, we try adding the redshift information of the spectroscopic galaxies in our sample as an additional feature in the SOM. 
This is an unconventional approach to an unsupervised machine learning method, since the norm is for the quantity being estimated to not be included among the SOM features; we indeed find that it is not beneficial, but we present our attempts for the sake of completeness.

Section \ref{sec:data} details the DES Year 3 data that we re-analyse with the proposed modifications to our redshift estimation method. Section \ref{sec:som} presents a summary of the Self-Organizing Map algorithm and the SOMPZ method for redshift estimation used by DES. Section \ref{sec:mod} presents the proposed modifications to the SOMPZ method used in DES Y3. Section \ref{sec:results:resdshifts} discusses the results of implementing these different modifications, and their impact on the redshift bins. Finally, Section \ref{sec:results:cosmo} shows the impact on cosmological parameter constraints.


\section{The Dark Energy Survey}\label{sec:data}

We summarize the samples used in this work in Table \ref{tab:catalogs}. These are the same ones used for redshift characterization of the weak lensing source galaxies in DES Year 3. The strategy employs Self-Organizing maps (SOMs), which we detail in Section \ref{sec:som}, and leverages the information present in three catalogs - wide, deep, and redshift - as well as Balrog injections:

\begin{table}
\centering
    \caption{Summary of the catalogs used in DES Year 3 for redshift estimation of the weak lensing source galaxies, including the area covered and the number of galaxies. The differing numerical precision reflects the distinct roles and sizes of the samples: the redshift and deep datasets are smaller but provide high-quality calibration information, while the wide sample covers almost the full survey footprint and contains a much larger number of galaxies whose redshifts we aim to infer by leveraging the information encoded in the other samples through the SOMs.}
    \begin{tabular}{ c|c|c } 
        \hline
         Sample & Area (sq. deg.) & Number of Galaxies \\
         \hline
         Wide & 4143 & 100,208,944 \\ 
         Deep & 5.88 & 2.8M \\ 
         Redshift & - &  57,000 \\ 
         Deep/Balrog & - &  2,417,437 \\
        \hline
    \end{tabular}
    \label{tab:catalogs}
\end{table}

\begin{itemize}
    \item[]{\bf Wide:} The weak lensing source catalog is described in detail in \cite*{DES:2020ekd}. After the applied selections $\hat{s}$ in magnitudes and colors, the {\it wide sample} is composed of $100,208,944$ galaxies, spread over $4143$ square degrees. DES has made flux measurements for all of these galaxies in the \textit{griz} bands of the electromagnetic spectrum (although the \textit{g}-band was not used in DES-Y3). 
    
    \item[]{\bf Deep:} The {\it deep sample} refers to the DES deep field galaxies, which have measured fluxes in additional bands \textit{ugrizJHK}. There are four deep-fields mapped in DES Y3, see \cite*{Hartley2021}, that added cover an area of $5.88$ square degrees. Notice that $Y$-band data in the deep fields had large offsets between the constituent exposures, and therefore could not be used.
    
    \item[]{\bf Redshift:}  A subset of the deep field galaxies have accurate redshifts obtained from a variety of external data sets. These include spectroscopic redshifts from multiple surveys, in addition to regions such as COSMOS2015 \citep*{2016ApJS..224...24L} where high-quality photometric redshifts are available (for full details of the sample and a list of external data included, see ~\citealt*{DES:2020ebm}). We call this set, containing approximately $57,000$ galaxies, the {\it redshift sample}. 
    
    \item[]{\bf Balrog:} In order to connect the information in our samples, we use Balrog injections. The Balrog software, developed by \cite*{Suchyta2016}, enables the creation of simulated galaxies, or Balrog injections, which are inserted into authentic images. These synthetic galaxies are designed based on the DES deep field photometry and are placed multiple times at various positions across the broader wide-field footprint, as specified in \cite*{Everett2022}. The resulting catalogue, called the \textit{deep/Balrog} sample, includes $2,417,437$ injection-realization pairs, each of which has both deep and wide photometric data. This sample is a crucial element of our redshift calibration inference technique. 
\end{itemize}

In what follows, we denote the wide data by $\hat{\pmb{x}}$ with covariance matrix $\hat{\Sigma}$ and the deep data by $\pmb{x}$ with covariance $\Sigma$, and the selection by $\hat{s}$, following the notation in \citet*{DES:2019bxr,DES:2020ebm}. 
The wide field data vector has three components, $\hat{\pmb{x}}=[r,i,z]$. For the deep fields, there are 3 infrared bands available -- J,H, and K -- and the DES $u,g$ bands are also used, such that $\pmb{x} =[u, g, r, i, z, J,H,K]$ has eight components in total. Since the redshift galaxies are a subset of the deep galaxies, they too have the 8 components $\pmb{x}$; Balrog galaxies typically have approximately 15 realizations $\hat{\pmb{x}}$ (corresponding to the number of wide field injections) for a single $\pmb{x}$ (corresponding to a single deep field galaxy).


\section{Self-organizing maps 
for photometric redshifts
}\label{sec:som}

In what follows, we review the SOM standard algorithm, and describe the SOMPZ method, i.e., how SOMs are used in practice for redshift estimation in DES-Y3.

\subsection{The SOM Algorithm}\label{sec:fiducial}

A Self-Organizing Map (SOM) is a type of Artificial Neural Network (ANN) that produces a discretized, lower dimensional, representation of the input space, while preserving its topology. Proposed by 
\citet*{SOM:1982}, it is an unsupervised Machine Learning method that uses soft competitive learning: the cells of the map (also known as nodes or neurons) compete to most closely resemble each training example until the best matching unit (BMU) is found, then the winner and its neighborhood are adapted.    

Consider a set of \textit{n} training samples, each with \textit{m} features, i.e., for each sample we have an input vector $\pmb{x}\in \mathbb{R}^{m}$. In our case, for instance, each sample is a single galaxy and the features are its fluxes (or colors or magnitudes) in $m$ bands. The SOM can be understood as collection of $C$ cells arranged in a \textit{l}-dimensional grid that has a specified topology. Each cell is associated with a weight vector $ \pmb{\omega}_k \in \mathbb{R}^{m}$, where $k = 1, \dotsb, C$. Both the input and weight vectors live in the input space, while the cells live in the output, or lattice, space. 

The training of a SOM is relatively straightforward. The weights are initialized to random or from data samples and the learning happens in three stages: Competition, cooperation, and weight adaptation.

\begin{itemize}
    \item Competition: at each step, a random sample of the training set is presented to the self-organizing map. The cell whose weight vector is the closest to the sample vector is identified as the best matching unit  (BMU): 
     \begin{equation}
        c_b = \argmin_k d(\pmb{x}, \pmb{\omega}_k).  
    \end{equation}
    The degree of "closeness" is determined by evaluating a distance metric $ d(\pmb{x}, \pmb{\omega}_k)$ between the sample and the cell in the map. Typically the Euclidian distance is used, but in \cite*{DES:2019bxr} the chi-square distance was chosen: 
    \begin{align}
        d^2(\pmb{x}, \pmb{\omega}_k) = (\pmb{x} -  \pmb{\omega}_k)^\top \pmb{\Sigma}^{-1}(\pmb{x} -  \pmb{\omega}_k),
        \label{eq:chisquare}
    \end{align}
    where $\pmb{\Sigma}$ is the covariance matrix for the training vector, $\pmb{x}$. The cell minimizing this distance is identified as the BMU, and the sample is then assigned to it.\\ 

    
    \item Cooperation: to preserve the topology of the input space, not only the BMU is identified and updated, but also its neighborhood. The neighborhood function $H_{b,k}(t)$ creates the connection between the input space and the cells in the map, responsible for the self-organizing property of the map. The size of the BMU's neighborhood decreases as a function of time steps, $t$. In addition, $H_{b,k}$ should decrease as the distance from the BMU increases. 
    It must also satisfy the properties that it is maximum in the winning cell $b$ and is symmetric about it. 
    A Gaussian neighborhood function attend those requirements:
    \begin{equation}
        H_{b,k}(t) = \exp[-D^2_{b,k}/\sigma^2(t)].
    \end{equation}

    The distance between the BMU, $c_b$, and any cell on the map, $c_k$, is the Euclidian distance in the \textit{l}-dimensional map:
    \begin{equation}
        D^2_{b,k} = \sum_{i=1}^{l}(c_{b,i}-c_{k,i})^2.
        \label{eq:euclidian}
    \end{equation}
    The width of the Gaussian kernel is given by 
    \begin{equation}
        \sigma(t) = \sigma_{s}^{1-t/t_{max}}.
    \end{equation}
    At the beginning of the training, $\sigma_s$ should be large enough that most of the map is initially affected. As the training progresses, the width shrinks until only the BMU and its closest neighbours are significantly affected by new data.\\   

    
    \item Weight adaptation: once the BMU is computed, we can calculate the updated value of the weight vectors for the $t + 1$-th iteration through the following relation:
    \begin{equation}
        \pmb{\omega}_k(t+1) = \pmb{\omega}_k(t) + a(t)H_{b,k}(t)[\pmb{x}(t)-\pmb{\omega}_k(t)],
    \end{equation}
    where $t$ is the current time step in training, $a(t)$ is the learning rate:
\begin{equation}
    a(t) = a_0^{t/t_{max}},
\end{equation}
    where $a_0 \in [0,1]$. In each iteration, this update function is applied to each of the cells in the map.
    
\end{itemize}

Different galaxy samples can be assigned to a trained SOM using the same methodology: for each object, the distance is computed, and the galaxy is assigned to the BMU. However, it is important to note that problems may arise when assigning a sample to a SOM that has been trained with a sample that is not representative of the color space of the samples to be assigned.

These steps describe the standard SOM algorithm, which has been applied for the purpose of redshift estimation in previous works (see e.g. \cite*{Masters2015,Speagle2019,DES:2019bxr}), including the DES-Y3 analyses \citep*{DES:2020ebm}.

\subsection{
Dark Energy Survey SOMPZ
}\label{sec:sompz}



In order to estimate the redshift distribution of the wide sample, we construct two SOMs: 
\begin{enumerate}
    \item Deep SOM: \vspace{-0.2cm}
        \begin{enumerate}
            \item Training sample: deep data
            \item Assignment sample: deep data (including the sub-sample with redshifts)
        \end{enumerate}    
    \item Wide SOM: \vspace{-0.2cm}
    \begin{enumerate}
            \item Training sample: wide data (or sufficiently large sub-sample thereof)
            \item Assignment sample: wide data
        \end{enumerate}
\end{enumerate}

\begin{figure*}
\centering
\includegraphics[width=2.\columnwidth]{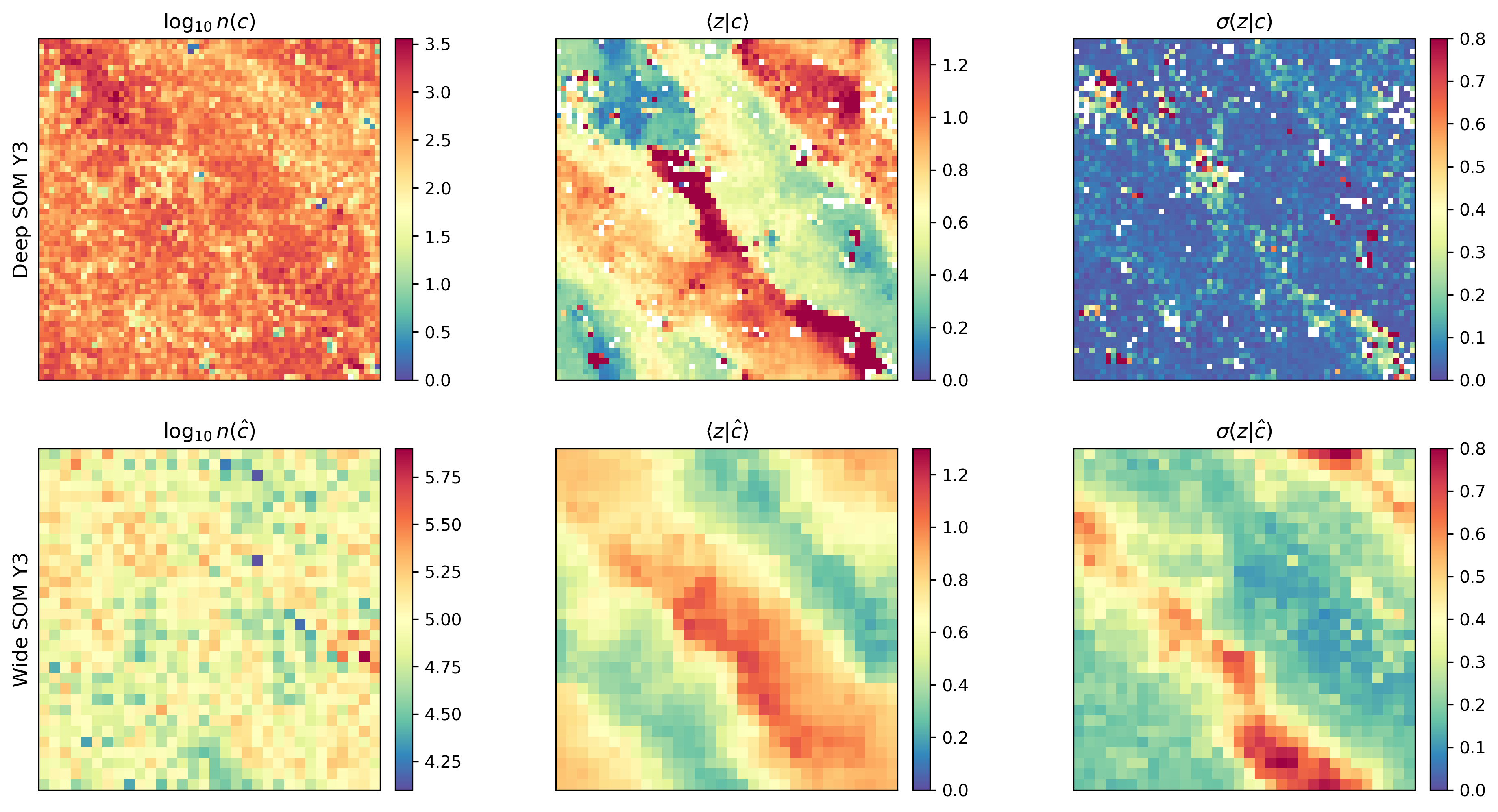}
\caption{
Visualization of the self-organizing maps constructed using the fiducial Y3 SOM algorithm described in Section \ref{sec:fiducial}. Top: Deep field self-organizing map
composed of 4096 cells. Bottom: Wide field self-organizing map composed of 1024 cells. The left-hand panels show the total number of galaxies assigned to
each SOM, the middle panels show the mean redshift for each cell, and the right panels show the standard deviation of the redshift distribution in each cell of
the map. 
The white cells found in the deep SOM are due to the lack of information in those regions of the color space, i.e., there are no galaxies
in the redshift sample that were assigned to those cells.}
\label{fig:somy3_deep_wide_std}
\end{figure*}

Figure \ref{fig:somy3_deep_wide_std} shows the number of objects, the redshift distribution and the standard deviation in the deep (top) and wide (bottom) SOMs used in the DES Y3 analysis. As we can see, and the name suggests, the deep SOM has deeper, and fainter, galaxies, going to higher $z$ values compared to the wide SOM. 

Each SOM cell acts as a way of discretizing the continuous color and color-magnitude spaces spanned by $\pmb{x}~(\hat{\pmb{x}})$ and $\Sigma~(\hat{\Sigma})$ into discrete categories $c~(\hat{c})$.
Therefore, the probability distribution function for the redshift of an ensemble of galaxies, conditioned on being observed in a particular cell $\hat{c}$, and on passing a selection function $\hat{s}$, can be written by marginalizing over the deep-field information:

\begin{equation}
     p(z|\hat{c},\hat{s}) =  \sum_c p(z|c,\hat{c},\hat{s}) p(c|\hat{c}, \hat{s}) .
\end{equation}

We then assign each cell $\hat{c}$ to a tomographic bin (see \cite*{DES:2020ebm} for the details on the assignment algorithm) and construct the $n(z)$ of each tomographic bin ($\hat{b}$) by summing over the cells belonging to the bin: 

\begin{align}
     n_{\hat b}(z)\equiv p(z|\hat{b},\hat{s}) &=  \sum_{\hat{c} \in \hat{b}} p(z|\hat{c},\hat{s}) p(\hat{c}|\hat{s},\hat{b})\\
     &= \sum_{\hat{c} \in \hat{b}} \sum_c p(z|c,\hat{c},\hat{s}) p(c|\hat{c}, \hat{s}) p(\hat{c}|\hat{s},\hat{b}).
\label{eq:pfinal}
\end{align}

In the equation above, each term is obtained from one of the galaxy samples we are using: 

\begin{enumerate}

    \item $p(z|c,\hat{c},\hat{s})$ is computed from the redshift sample subset of the deep sample, which contains spectroscopic redshifts, deep photometry, and wide-field Balrog realisations. It tells us the probability of getting a redshift $z$, given the deep cell $c$, the wide cell $\hat{c}$, and the selection $\hat{s}$.

    \item $p(c|\hat{c}, \hat{s})$ is computed from the Balrog injections of the entire deep sample. It tells us the probability of ending up in the deep cell $c$, given the wide cell $\hat{c}$ and the selection $\hat{s}$. We call this term the transfer function, because it connects the deep and wide cells. It is computed from Balrog realisations, because it requires both wide-field and deep-field photometry to be available.
    
    \item $p(\hat{c}|\hat{s},\hat{b})$ is computed from the wide sample.
    It tells us the probability that a galaxy in bin $\hat b$ is in the wide SOM cell $\hat c$. Therefore, cells with very few galaxies in them are down-weighted when determining the redshift distribution of the bin\footnote{In order to address the concern relative to over- and under-weighting galaxies when using raw fluxes, luptitudes were used for the input vectors.}.

\end{enumerate}

Assuming that the $p(z)$ in the deep cells (with high quality photometry) do not depend on the wide (noisy) photometry of those galaxies, we can remove the conditions on $\hat{c}$ and $\hat{b}$ in the first and last terms of Equation \ref{eq:pfinal}, and approximate it to

\begin{align}
     p(z|\hat{b},\hat{s}) 
     &\approx \sum_{\hat{c} \in \hat{b}} \sum_c p(z|c,\hat{s}) p(c|\hat{c}, \hat{s}) p(\hat{c}|\hat{s}).
\end{align}

The transfer function, $p(c|\hat{c}, \hat{s})$, connecting the deep and wide samples, is computed from Balrog realisations, not the full wide galaxy sample. Re-writing it as 

\begin{equation}
    p(c|\hat{c}, \hat{s}) = \frac{p(c ,\hat{c}|\hat{s})}{p(\hat{c}|\hat{s})},
\end{equation}
and replacing it in the equation above, we can write each term highlighting the sample from which it is obtained

\begin{equation}
    p(z|\hat{b},\hat{s}) \approx \sum_{\hat{c} \in \hat{b}} \sum_c \underbrace{p(z|c,\hat{s})}_{\text{Redshift}} \underbrace{p(c|\hat{s})}_{\text{Deep}} \underbrace{\frac{p(c,\hat{c}|\hat{s})}{p(c|\hat{s}) p(\hat{c}|\hat{s})}}_{\text{Balrog}} \underbrace{ p(\hat{c}|\hat{s})}_{\text{Wide}}. 
    \label{eq:nzsamples}
\end{equation}

\vspace{0.25cm}

We would like to emphasize that solving Equation \ref{eq:nzsamples} is not the final result for the Y3 $n(z)$'s, as two other pieces of information were added in: clustering redshifts and shear ratios (see \cite*{y3-sourcewz}, \cite*{y3-shearratio}). However, this is the main result to which we are interested in comparing in this work. In what follows we will compare this \textit{fiducial Y3} $n(z)$ to the one obtained by each SOM modification proposed in this paper.

\section{
Testing improved SOM methodology
}\label{sec:mod}

In this section we describe the three modifications to the standard method,  and assess the impact on the DES Y3 redshift distributions: replacing the SOM algorithm used in Year 3 (see \citealt*{DES:2019bxr}) by the one proposed in \citet*{Sanchez2020}; including an extra band (\textit{g}-band), even though it has low SNR; including redshifts, when available, as an additional feature to train and assign galaxies to the SOM.

\subsection{SOM for faint galaxies - SOMF}\label{sec:somf}

\begin{figure*}
\centering
\includegraphics[width=2.\columnwidth]{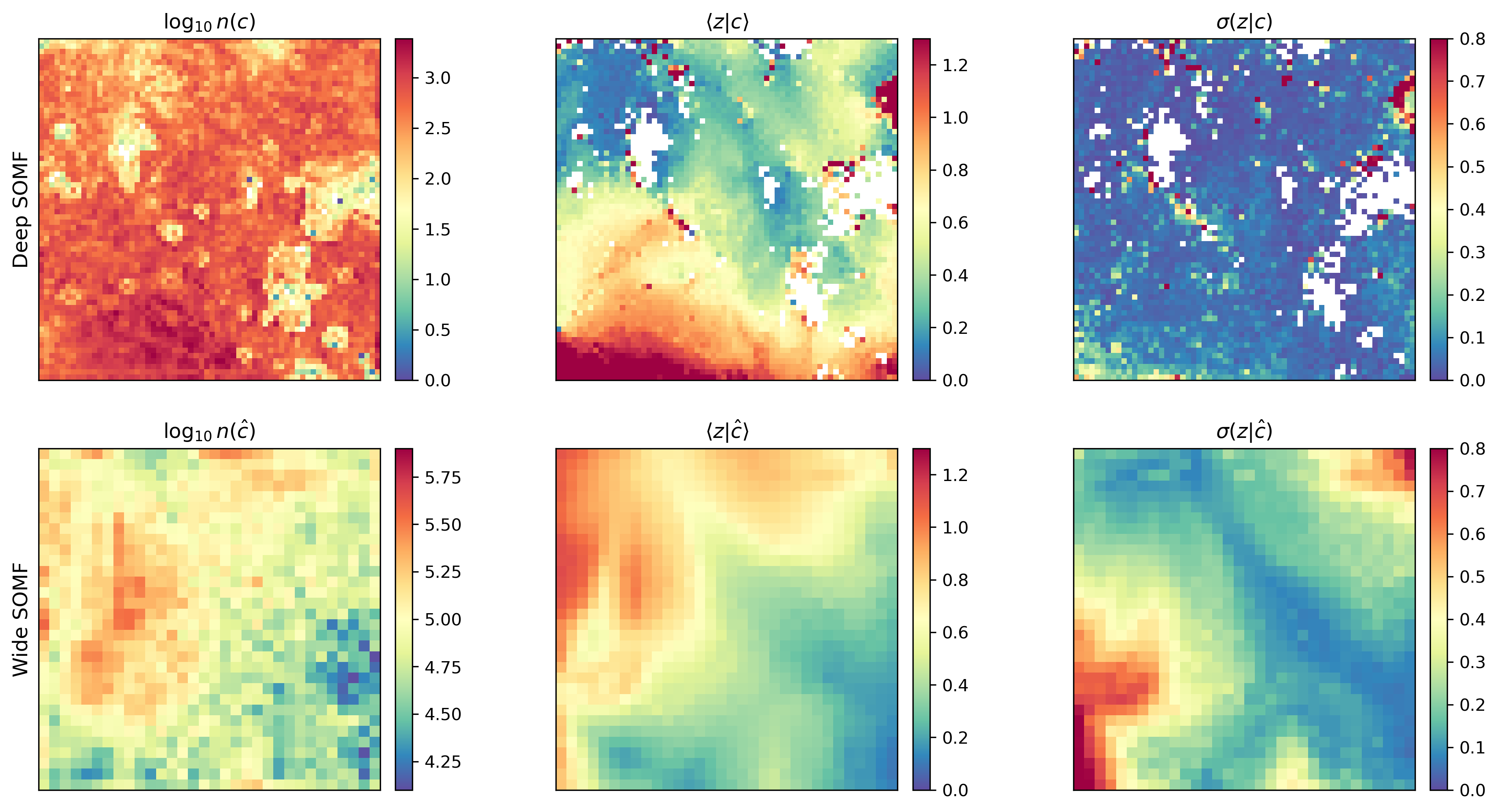}
\caption{Visualization of the self-organizing maps constructed using the SOMF algorithm described in Section \ref{sec:somf}. Top: Deep field self-organizing map composed of $4096$ cells. Bottom: Wide field self-organizing map composed of $1024$ cells. The left-hand panels show the total number of galaxies assigned to each SOM, the middle panels show the mean redshift for each cell, and the right panels show the standard deviation of the redshift distribution in each cell of the map. The white cells found in the deep SOM are due to the lack of information in those regions of the color space, i.e., there are no galaxies
in the redshift sample that were assigned to those cells.}
\label{fig:somf_deep_wide_std}
\end{figure*}

A characteristic of the majority of machine learning methods, self-organizing maps included, is the assumption that the training data is ideal, i.e., does not contain errors. This assumption is not true in general, especially when working with empirical data. 
This point is addressed for the case of SOMs in \citet*{Sanchez2020}, where the authors propose a modification of the standard SOM algorithm that accounts for measurement uncertainties in the training set, with the problem of faint galaxies in mind. 
The basic idea is to take the errors into account such that, examples with larger measurement uncertainties will result in less change to the weights than examples with smaller uncertainties. 
The main modifications to the standard algorithm consist in redefining the distance measure between a training sample and a cell on the map, and the training shift through which the weights are updated. 

In addition, the sample features ($\pmb{x}$) and cell weights ($\pmb{\omega}_k$) are converted into units of signal-to-noise ratio (SNR), specifying a maximum for the sample SNR as a means of softening the specificity of the cells:
\begin{align}
    s_{ib} &\equiv \max\left(\Sigma_{ib}, \frac{x_{ib}}{\text{SNR}_{\max}}\right) \\
    \nu_{ib} &\equiv \frac{x_{ib}}{s_{ib}} \\
    \nu_{cb} &\equiv \frac{\omega_{cb}}{s_{ib}}
\end{align}
Here, the index $i$ represents each individual galaxy, the index $c$ represents each cell, and the quantities $\nu_{ib}$ and $\nu_{cb}$ are the galaxy fluxes and cell weights (indexed by the photometric band b). This helps to ensure that the SOM does not overreact to outliers or extreme values in the data, and makes the SOM more robust or, less sensitive, to variations in the SNR of the data samples.

The SOM algorithm presented in Section \ref{sec:som} uses the chi-square distance, defined in Equation \ref{eq:chisquare}, as the metric between the sample and the cell in the map. In \citet*{Sanchez2020}, however, the authors define: 

\begin{equation}
    d(\pmb{x},\pmb{\Sigma},\pmb{\omega}_k) = \inf_s\left[\tilde{d}(\pmb{x},\pmb{\Sigma},e^s\pmb{\omega}_k)+\frac{s^2}{\sigma^2_s}\right],
    \label{eq:dist_fuzzy}
\end{equation}
where
\begin{align}
    \tilde{d}(\pmb{x},\pmb{\Sigma},\pmb{\omega}_k) = &\sum_b\left[\frac{\arcsinh \nu_{cb}+W_{ib}\log 2\nu_{cb}}{1+W_{ib}}-\arcsinh \nu_{ib}\right]^2\nonumber \\ 
    &\times(1+\nu^2_{ib}),
\end{align}
approaches the Euclidean distance in log-flux at high SNR, and is also Euclidean in linear flux at low SNR, while weighting each band by its SNR (up to a maximum). As a result this metric is better suited to the wide dynamic range of galaxy fluxes. The weighting function is defined as 
\begin{equation}
    W_{ib} = e^{2(\nu_{ib}-4)},
\end{equation}
such that it is possible to transition from the high- to low-SNR regimes.
Equation \ref{eq:dist_fuzzy} includes an overall scale constant $e^s$ which allows the cells to be “fuzzy” in overall flux level. 
As pointed out in \citet*{Sanchez2020}, there is no natural periodicity in the feature space of galaxy colors and magnitudes. 
Therefore, the assumption of periodic boundary conditions, usual to the standard algorithm, is not adopted here.

For SOMF, the input vector consists of fluxes. This choice addresses the challenge of color stability in low signal-to-noise (S/N) scenarios, as lupti-colors of faint galaxies tend not to be very reliable. Furthermore, the distance estimator in Equation \ref{eq:dist_fuzzy} ensures that, for galaxies with significant noise in certain bands, the SOM assignment does not rely on noisy color information, while still finely distinguishing galaxy types through their color profiles in high S/N situations.

\subsubsection{
Application to DES Y3
}
\label{sec:application_des_y3}
We test the impact of using this modified SOM methodology with the DES Y3 data.
Figure \ref{fig:somf_deep_wide_std} shows a deep (top) and a wide (bottom) SOM constructed using the DES Y3 data described in Section \ref{sec:data} and this modified SOM algorithm. 
The left panels show the number of objects distributed in each of the two SOMs.
The smooth behavior of redshift across the SOM, as seen in the middle panels, shows that the variation of redshift in both the 8-band and 3-band space topology, shown in Figures \ref{fig:wide_somf_y3} and \ref{fig:deep_somf_y3}, are reasonably well traced by the 2D SOM.
The right-most panels show the standard deviation of the redshift distribution in each SOM cell. 

We emphasize that the smoothness present in the middle panel of the SOMs is evidence that the redshift space is being well mapped by the flux space. 
We see that the transitions between low- and high-redshifts do not happen abruptly, in general, as we expect in a successful compression of this high dimensional space. 
The white cells in the top panels represents cells that ended up without galaxies from the redshift sample. Therefore we could not estimate the redshift distribution in those cells, and they were not used. 
Comparing Figure \ref{fig:somy3_deep_wide_std} and Figure \ref{fig:somf_deep_wide_std} we see that, even though the fiducial Y3 Deep SOM also had white cells, there were fewer than using the SOMF algorithm.
This is not an issue, given that the amount of information in the transfer function is still the same, however it further emphasizes that the two SOM algorithms group galaxies from the same catalog in a different way. 
In a recent paper, \cite*{sanchez_alarcon2022} argue that the reason for this could be related to the SOMF algorithm being better at anomaly detection, and this difference could come from strange, misdetected objects in our catalog. Notice that the map initialization when running SOM and SOMF is not the same, i.e., it is not possible to do a cell-to-cell comparison between the SOMs in Figure \ref{fig:somy3_deep_wide_std} and Figure \ref{fig:somf_deep_wide_std}. 

\begin{figure}
\centering
\includegraphics[width=\linewidth]{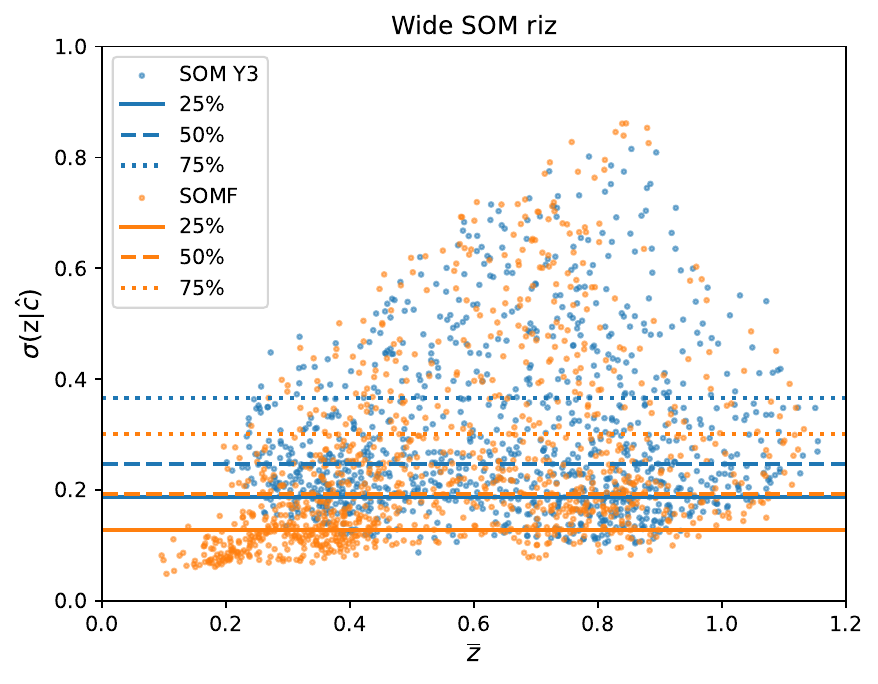}
\caption{Standard deviation $\sigma(z|\hat{c})$ of the redshift distribution in each wide SOM cell, versus the mean redshift $\overline{z}$ of each cell, for the standard Y3 SOM (blue), and the SOMF (orange). The horizontal lines represent the $25$ (solid), $50$ (dashed) and $75$ (dotted) percentiles of $\sigma(z|\hat{c})$. We observe that the SOMF presents an overall reduction in the uncertainty per wide cell.}
\label{fig:std_somf_y3som}
\end{figure}

An additional feature to notice is that comparing the lower-right panels in both figures, showing the standard deviation of the redshift distribution in each wide SOM cell, the SOMF exhibits a significantly larger fraction of blue, i.e. low-standard deviation, cells.
Figure \ref{fig:std_somf_y3som} compares the standard deviation of the redshift distribution in each cell of the wide SOM  $\sigma(z|\hat{c})$ for the SOMF and the Y3 SOM, showing an overall improvement when using the SOMF algorithm. The horizontal lines represent the $25$ (solid), $50$ (dashed) and $75$ (dotted) percentiles of $\sigma(z|\hat{c})$. We can observe that for the SOMF we have about $50\%$ of the wide cells with $\sigma(z|\hat{c}) < 0.2$, while that is true for only about $25\%$ of the Y3 SOM wide cells. This represents a $\sim 20\%$ reduction in the median of the standard deviation for SOMF when compared to the Y3 SOM.

\subsection{Regaining blue bands for redshift estimation - \textit{griz}\label{sec:somg}}
Although measured in wide field photometry, the \textit{g}-band did not have an accurate enough point spread function to measure the shapes of galaxies. 
In particular, the \textit{g}-band rho statistics (see \cite*{jarvis2020} Figure 13) were considered unacceptably large, which led to the exclusion of \textit{g}-band data from the Y3 weak lensing analysis.
Here we perform the exercise of including \textit{g}-band information (obtained from \citealt*{y3-gold}, and matched to \citealt*{DES:2020ekd}), in addition to \textit{r, i, z} bands, to create and assign galaxies to the wide SOM.  
Notice that, since the Metacal convolution and deconvolution \citep*{DES:2020ekd} could not be carried out, we do not have shape measurements with \textit{g}-band for Y3, therefore this exercise is purely at the photometric redshift level.  

\subsubsection{Application to DES Y3}

The samples used here are exactly the same as used in obtaining the fiducial Y3 weak lensing redshift measurements (see Section \ref{sec:data}), the only difference is the inclusion of the \textit{g}-band in training the wide SOM and, therefore, including \textit{g}-band fluxes when assigning the wide and balrog samples to the wide SOM. Our purpose is to quantify the improvement in our redshift constraints, in the hypothetical case that the \textit{g}-band measurements had been considered good enough to use in the DES Y3. This is particularly timely because we expect that for the Y6 analysis the \textit{g}-band PSF solution will be sufficiently improved by the addition of color dependence in the PSF model, allowing it to be used for the weak lensing analysis \citep*{jarvis2020}.

\begin{figure*}
\centering
\includegraphics[width=2.\columnwidth]{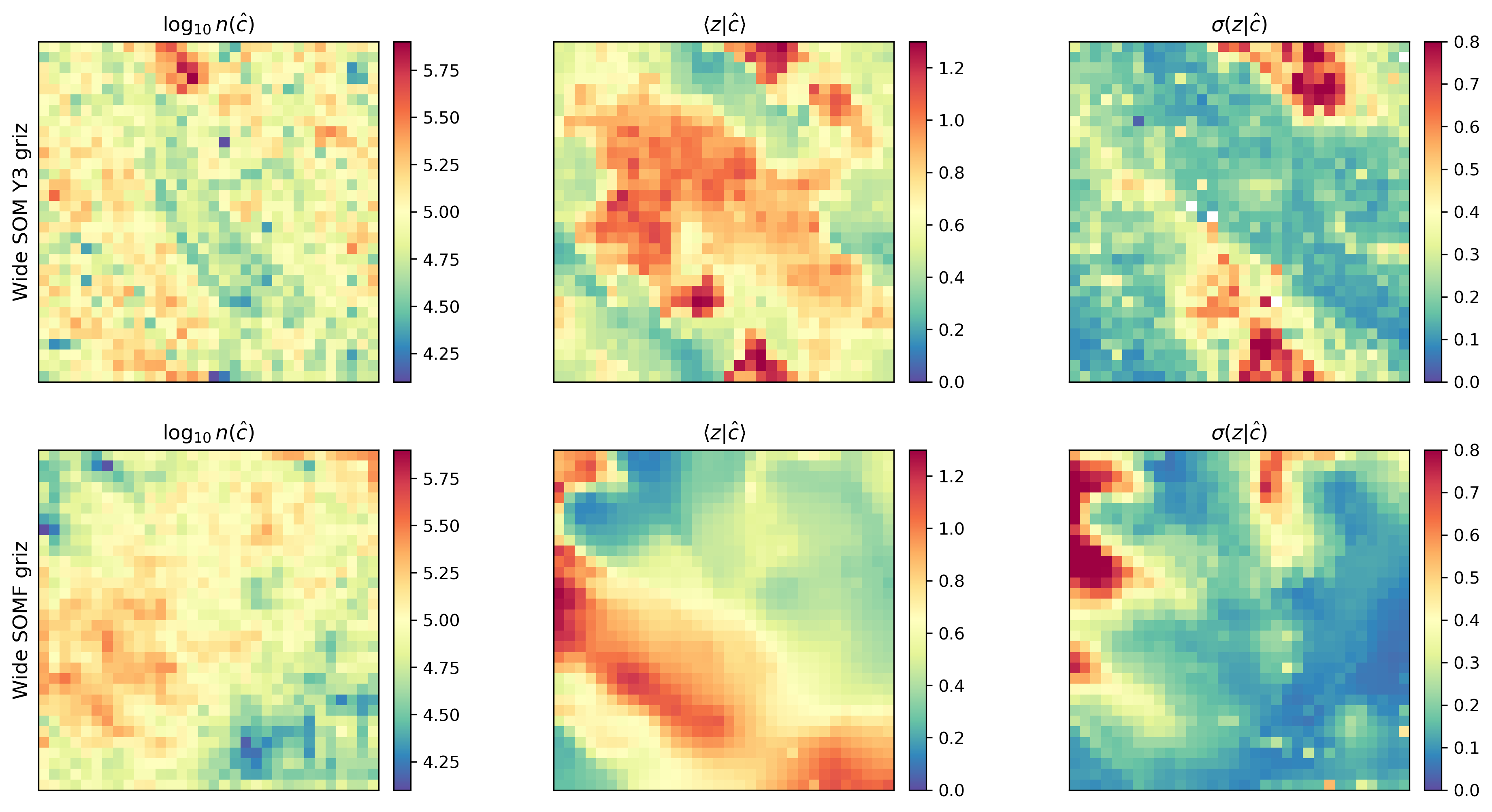}
\caption{Visualization of the self-organizing maps constructed adding the \textit{g}-band to train and assign the wide data, as described in Section \ref{sec:somg}. Top: Wide field self-organizing map obtained using the DES Y3 SOM algorithm, but adding the wide sample \textit{g}-band information. As discussed in Section \ref{sec:application_des_y3}, the smoother map of mean redshift in wide SOM cells when using the SOMF reflects an improved compression of the color-space. Bottom: Wide field self-organizing map obtained using the SOMF algorithm, but adding the wide sample \textit{g}-band information. 
Notice that the introduction of the wide sample \textit{g}-band does not affect the deep SOM, which already includes the deep sample \textit{g}-band, and therefore it is the same as in Figures \ref{fig:somy3_deep_wide_std} and \ref{fig:somf_deep_wide_std} and omitted here. The left-hand panels show the total number of galaxies assigned to each SOM, the middle panels show the mean redshift for each cell, and the right panels show the standard deviation of the redshift distribution in each cell of the map.}
\label{fig:deep_somg}
\end{figure*}

We test the impact of the addition of the \textit{g}-band using the fiducial Y3 SOM algorithm (see Section \ref{sec:som}), and the modified SOM described in Section \ref{sec:mod}. Adding the \textit{g}-band impacts only the wide SOM part of the SOMPZ method. Figure \ref{fig:deep_somg} shows wide SOM constructed using the fiducial Y3 SOM described in Section \ref{sec:som} (top), and the one described in Section \ref{sec:mod} (bottom), adding the \textit{g}-band information to the train the wide SOM and assigning data to it.  
The left panels show the number of objects distributed in each of the two SOMs. The middle panels show the mean redshift for each cell. Notice that the inclusion of the \textit{g}-band creates more cells at higher redshifts in the wide SOM, when compared to either the Y3 fiducial SOM or the SOMF, indicating that the additional \textit{g}-band information results in fewer cells mistakenly assigned to low redshifts.
The right-most panels show the standard deviation of the redshift distribution in each SOM cell, which is compatible with what we see for the Y3 fiducial SOM and the SOMF.

\subsection{Including redshift}\label{sec:somz}

The original deep SOM is trained using deep field galaxies and assigned using redshift sample galaxies and deep field galaxies. This enables us to infer the deep field galaxy $p(z|c,\hat{s})$ distribution from the redshift sample galaxies for each cell. The redshift itself is {\it not} used as a feature; only the photometric fluxes in each band are in the feature vector. Here, we investigate the impact of including redshifts of galaxies (when available) as an extra feature, such that for each sample we have an input vector $\pmb{x}$:

\begin{equation}
   \pmb{x} = \left[u,g,r,i,z,J,H,K,{\rm redshift} \right],
\end{equation}
containing the 8 fiducial bands plus the redshift information. For deep field galaxies that do not have redshift information, their cells are still determined based only on their fluxes. 

To quantify the contribution of redshift in the training and assigning process, we use a weighting factor $\lambda$ to modulate how much it contributes relative to the photometric bands. When $\lambda=1$, the redshift information is normalized to have the same contribution as a flux. We also consider $\lambda=0.1$ and 0.05, in which cases redshift contributes only $10\%$ and $5\%$ that of a flux, so that the redshift information plays a smaller role in constructing the SOM and assigning galaxies to cells.

The methodology and samples used here are exactly the same as used in obtaining the fiducial Y3 weak lensing redshift measurements (see Sections \ref{sec:data} and \ref{sec:som}). 
The only difference is the inclusion of the redshift of galaxies as an extra feature in training the deep SOM. 
We test two possibilities (i) including redshift information both in training and assigning galaxies to the SOM and (ii) in the SOM training process only. The results relative to these tests are shown in Figures \ref{fig:som addz} and \ref{fig:som tzaf}.

\section{Redshift Bins and Bin Overlap}\label{sec:results:resdshifts}

Having well-defined redshift bins is essential for weak lensing analysis, in order to ensure accurate and unbiased measurements of the gravitational lensing effect. Minimizing the overlap of redshift distributions reduces the contamination of signals between bins, which is crucial to probe the lensing signal as a function of source redshift, control systematic errors, and enable precise cosmological parameter constraints. For instance, having narrower/less-overlapping bins reduces the Intrinsic Alignment (IA) signal (specially the II component in cross-bin pairs), reducing the impact of any potential mismodelling. In addition, having finer definition along the line of sight helps constrain the evolution of structure, so it is also better from a purely cosmological perspective.

Using the DES Y3 data, described in Section \ref{sec:data}, we assessed the ability of the modifications to the fiducial Y3 SOMPZ method (see Section \ref{sec:som} and \citealt*{DES:2020ebm}), detailed in Section \ref{sec:mod}. In particular, replacing the fiducial Y3 SOM by the SOMF and including the \textit{g}-band information improves our redshift constraining power and reduces the bin overlap. We detail our findings regarding those modifications in what follows. The third possibility that we described in Section \ref{sec:somz}, adding the redshift information, when available, to train and assign galaxies to the SOM, does not provide any improvements in comparison to the Y3 results. Therefore, we move our findings on that to Appendix \ref{sec:appendixA}.

\subsection{\textit{N(z)} distributions}\label{sec:results:nzs}

\begin{table*}
    \begin{tabular}{l c c c c}
     & Bin 0 & Bin 1 & Bin 2 & Bin 3 \tabularnewline
     $z^{PZ}$ range &  0.0—0.358 & 0.358—0.631 & 0.631—0.872 & 0.872—2.0\tabularnewline
    \hline 
    \hline
    $\langle z \rangle$ Y3 SOM & 0.335 & 0.518 & 0.750 & 0.936 \tabularnewline
    $\langle z \rangle$ SOMF & 0.327 & 0.510 & 0.735 & 0.928 \tabularnewline
    
    \hline
    $\langle z \rangle$ Y3 SOM \textit{griz} & 0.328 & 0.473 & 0.729 & 0.968 \tabularnewline
    $\langle z \rangle$ SOMF \textit{griz} & 0.312 & 0.467 & 0.725 & 0.976 \tabularnewline
    \hline 
    \hline
    \bf{Uncertainty*} &  &  & \tabularnewline
    \hline
    \hline
    Shot Noise $\&$ Sample Variance & 0.006 & 0.005 & 0.004 & 0.006 \tabularnewline
    Redshift Sample Uncertainty & 0.003 & 0.004 & 0.006 & 0.006 \tabularnewline
    Balrog Uncertainty & < 0.001 & < 0.001 & < 0.001 & < 0.001 \tabularnewline
    Photometric Calibration Uncertainty & 0.010 & 0.005 & 0.002 & 0.002 \tabularnewline
    Inherent SOMPZ Method Uncertainty  & 0.003 & 0.003 & 0.003 & 0.003 \tabularnewline
    Combined Uncertainty: SOMPZ (from 3sDir) & 0.012 & 0.008 & 0.006 & 0.009 \tabularnewline
    \hline 
    \hline 
    \textbf{*} We refer to \citet*{DES:2020ebm} for the definition of each uncertainty.
    \end{tabular}
    \caption{Values of and approximate error contributions to the mean redshift of each tomographic bin. Given that the the only difference between the redshift distributions estimated using the Y3 SOM and the SOMF comes from the SOM recipe (all the samples are the same in both cases), we can safely assume that the uncertainties due to Shot Noise $\&$ Sample Variance, Redshift Sample, Balrog  and Photometric Calibration are exactly the same ones estimated for DES Y3 \citep*{DES:2020ebm}. The only uncertainty affected by the change in our method is the inherent SOMPZ Method uncertainty. Figure \ref{fig:std_somf_y3som} suggests that uncertainty to be even smaller for the SOMF, therefore we decided to not re-compute the SOMPZ uncertainty, and assume its upper bound to be the same as the Y3 SOM.} 
    \label{table:errors}
\end{table*}

\begin{figure}
\centering
\includegraphics[width=\linewidth]{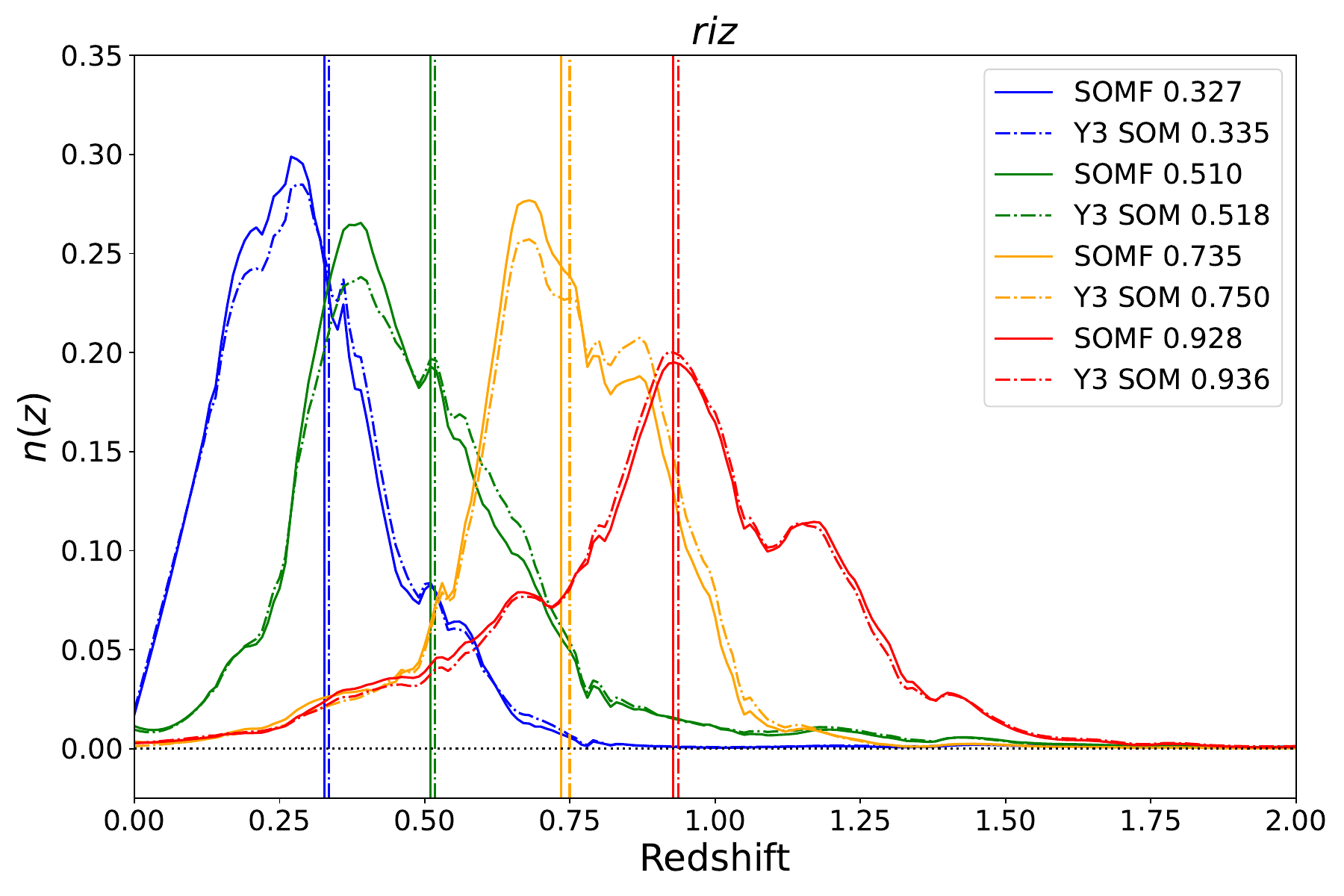}
\caption{Photometric redshift distribution obtained from the \textit{riz} bands, using the Y3 SOM (dot-dashed line) and the SOMF  algorithm (filled line). The two methods show good agreement regarding the shape of each bin, and their mean redshifts values, shown in the legend on the top right. The SOMF method, however, presents better defined bins.}
\label{fig:nz_somf}
\end{figure}

In Figure \ref{fig:nz_somf}, we compare the result of applying the redshift schema described in Section \ref{sec:sompz}, in particular the solution of Equation \ref{eq:pfinal}, using both the Y3 SOM and the SOMF. We can see that the two methods agree very well, both in mean redshift and shape of the $n(z)$.
In particular, once we apply the uncertainties due to each component of the method, shown in Table \ref{table:errors}, all bins agree well inside the uncertainty level, with the exception of bin 2 that is slightly off the uncertainty bound. Notice that we can safely assume that the uncertainties due to Shot Noise and Sample Variance (caused by the size and area of the deep fields), irreducible biases in the Redshift Samples, the use of Balrog, and variations in the Photometric Calibration across deep fields are exactly the same as those estimated for DES Y3 (see \citealt*{DES:2020ebm} for details on each uncertainty and how they were estimated). The inherent SOMPZ Method uncertainty is the only one affected by the change of method, but given the good agreement in mean redshift and shape of the distributions, we decided to not recompute it, and assume it is the same as for Y3 as well. This is further justified by the fact the contribution of the SOMPZ uncertainty to the total error budget per tomographic bin was minor. Notice that the SOMF method produced bins seem slightly better defined, with higher peaks, even though the two distributions follow each other very closely.

\begin{figure}
\centering
\includegraphics[width=\linewidth]{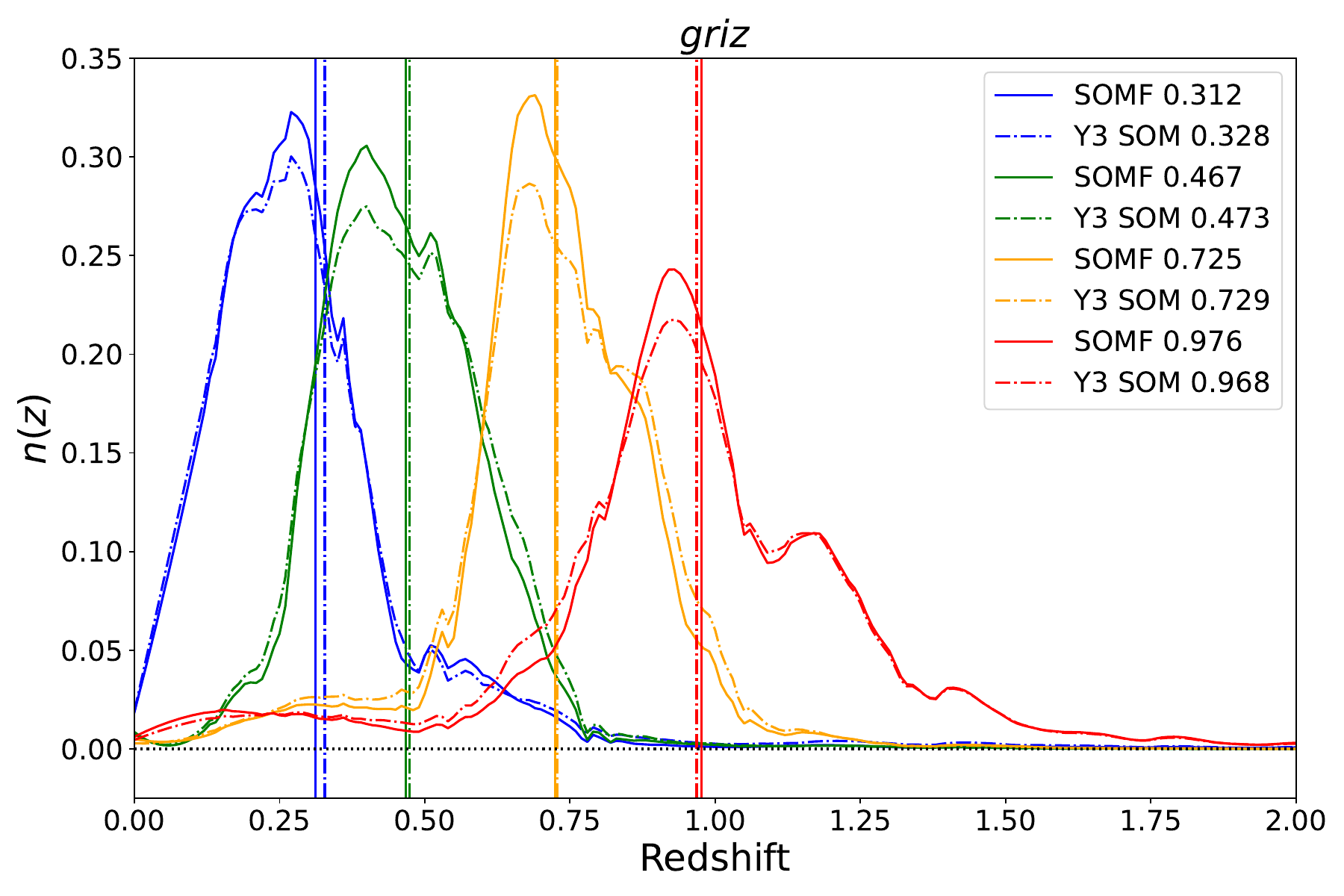}
\caption{Photometric redshift distribution obtained from the \textit{griz} bands, using the Y3 SOM (dot-dashed line) and the SOMF  algorithm (filled line). The two methods show good agreement regarding the shape of each bin, and their mean redshifts values, shown in the legend on the top right. However, the addition of the \textit{g}-band further emphasizes the ability of SOMF to produce better defined bins.}
\label{fig:nz_g}
\end{figure}

Figure \ref{fig:nz_g} shows a similar comparison, but now including the \textit{g}-band information, for for the fiducial Y3 SOM and the SOMF. We can see that two SOM algorithms again agree very well, both in mean redshifts and shape of the $n(z)$. The mean redshift in each bin agree within the uncertainty level in Table \ref{table:errors}.
The difference in the peak heights, is even more pronounced now with the addition of the \textit{g}-band information, showing that the SOMF algorithm leverages the \textit{g}-band information to get even better defined redshift bins. 
Notice that the means and shapes of the distributions in Figure \ref{fig:nz_somf} and Figure \ref{fig:nz_g} differ from each other, which is a expected consequence of the addition of the \textit{g}-band information.

\subsection{Bin overlap}\label{sec:results:bin}

\begin{figure}
\centering
\includegraphics[width=\linewidth]{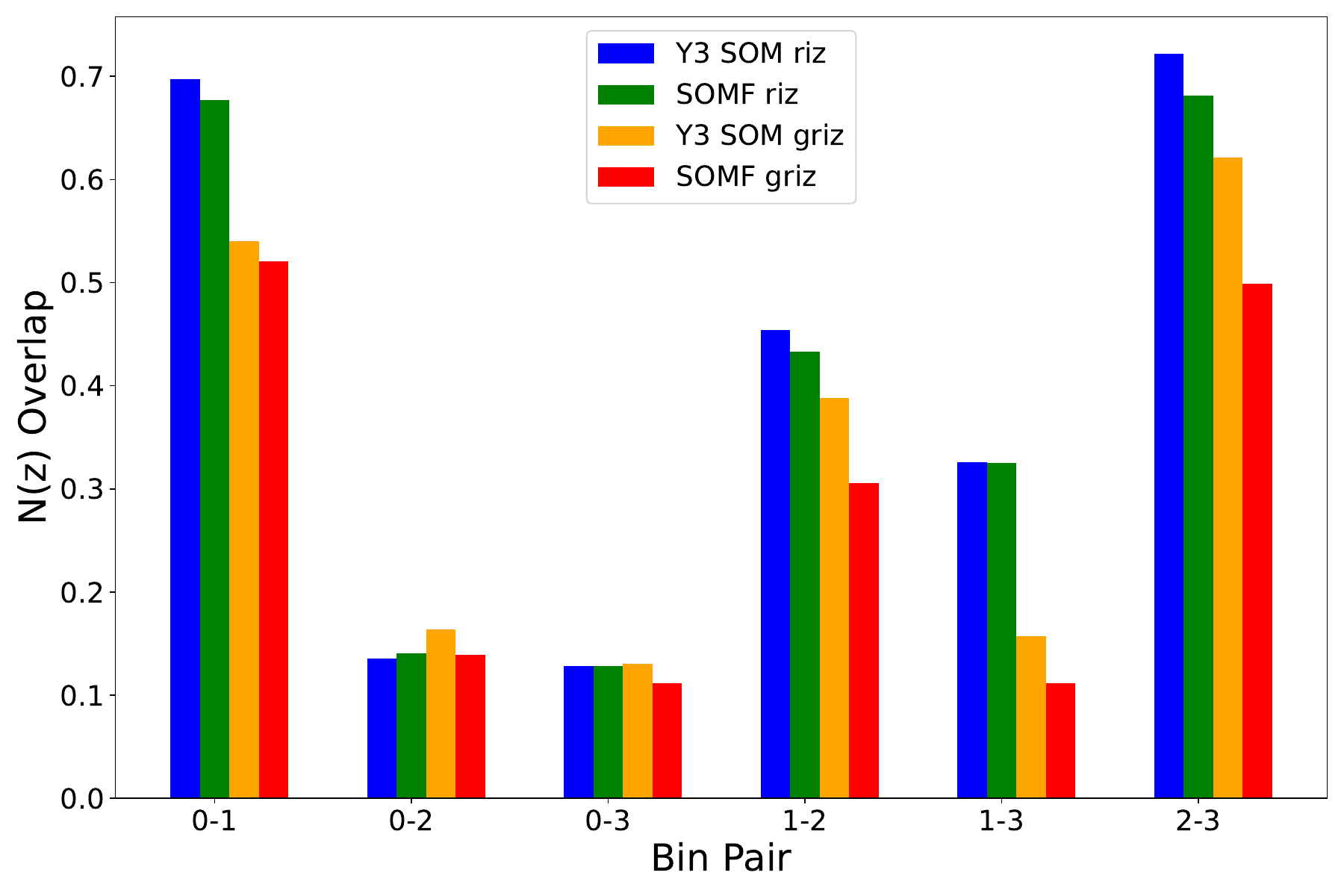}
\caption{Redshfit bin overlap between bins $0-1$, $0-2$, $0-3$, $1-2$, $1-3$ and $2-3$ for each SOM recipe. The Y3 SOM \textit{riz} is shown in blue, the SOMF \textit{riz} in green, the Y3 SOM \textit{griz} in yellow and the SOMF \textit{griz} in red. We can see that all proposed modifications reduce the bin overlap with respect to the fiducial Y3 SOM using \textit{riz} bands, with the best result obtained for SOMF \textit{griz}.}
\label{fig:overlap}
\vspace{-1cm}
\end{figure}

We aim for minimal overlap between bins, indicating distinct redshift ranges that have been well separated. This results in a higher likelihood that a galaxy is correctly assigned to its designated bin rather than to a neighboring one.
Figure \ref{fig:overlap} compares the amount of bin overlap obtained with each method. The amount of bin overlap when using the Y3 SOM and the \textit{riz} bands is shown in blue, the SOMF with \textit{riz} bands is shown in green, Y3 SOM with \textit{griz} bands in yellow, and the SOMF with \textit{griz} bands in red. We can immediately see that the Y3 SOM \textit{riz} presents the highest overlap among all methods, and the greatest reduction in bin overlap is obtained when we combine the SOMF recipe and the \textit{griz} bands. 

The bin overlap is calculated by first computing an overlap matrix and then normalizing it. The overlap matrix is essentially a matrix of dot products between all pairs of bins in the $n(z)$ distribution. Mathematically, this can be represented as
\begin{equation}
G = X X^\top,
\end{equation}
where 
$X$ is an $n \times m$ matrix representing  $n(z)$, with $n=4$ bins (assuming 
$X$ has $4$ rows and $m$ columns). This matrix 
$G$ is akin to a Gram matrix in linear algebra, which contains the dot products of vectors in a set. 
We then normalize the overlap matrix by dividing it element-wise by the product of the square roots of its diagonal elements. In other words, we compute a normalization matrix 
$D$ where each element $D_{ij} = \sqrt{G_{ii}G_{jj}}$. The final normalized overlap matrix is obtained by
\begin{equation}
G_{\text{normalized}} = \frac{G}{D}.
\end{equation}
This step adjusts the overlap values based on the scale of the data, which is crucial to ensure that the results are not skewed by differences in variance among datasets. The normalized Gram matrix is used as a measure of similarity in various applications, including clustering methods in machine learning (see e.g. \citealt*{gram}).

The numerical values corresponding to the bin overlap between bin pairs in shown in Table \ref{tab:overlap}, where we also show the percent decrease in bin overlap relative to the Y3 SOM \textit{riz}. The SOMF \textit{riz}, Y3 SOM \textit{riz}, SOMF \textit{griz} present decreasing amount in overlap, having a reduction of 
3$\%$, 23$\%$, and 25$\%$ respectively for bins 0-1; 5$\%$, 14$\%$, and 33$\%$ for bins 1-2; 0$\%$, 52$\%$, and 66$\%$ for bins 1-3; 6$\%$, 14$\%$, and 31$\%$ for bins 2-3. In the case of the overlap between bins 0-2 and 0-3, the amount of overlap is already small when compared to the other bin pairs. For those two pairs all methods yield similar results, with the Y3 SOM \textit{riz} having the best performance by a few percent for bins 0-2 --- 
where the slight increase in overlap between bins 0 and 2 when including the \textit{g}-band likely arises from residual degeneracies in galaxy spectral energy distributions at the corresponding redshifts --- 
and the SOMF \textit{griz} again having the best performance for bins 0-3.

In summary, using the same data and same methodology, described in Section \ref{sec:sompz}, we can reduce the amount of bin overlap in our wide sample just by replacing the fiducial Y3 SOM algorithm used in the Y3 analysis (see \citealt*{DES:2020ebm}, \citealt*{DES:2019bxr}), described in Section \ref{sec:som}, by the SOMF algorithm (see \citealt*{Sanchez2020}) described in Section \ref{sec:mod}, and adding the \textit{g}-band information. By adding the \textit{g}-band, the overlap between redshift bins undergoes significant improvement. The fiducial Y3 SOM already shows significant reduction in bin overlap when the \textit{g}-band is added, but it is by combining the SOMF with the \textit{g}-band information that we obtain a substantial reduction in bin overlap and the best defined redshift bins. In the next section, we show the impact of this reduction in bin overlap on the cosmological parameters.

\begin{table}
    \centering

    \begin{tabular}{lcccc}
    \multicolumn{4}{c}{\hspace{2.7cm} Method}\tabularnewline
              & Y3 SOM  & SOMF  &Y3 SOM   & SOMF  \tabularnewline
    Bin Pairs &  \textit{riz}  & \textit{riz} & \textit{griz}  & \textit{griz} \tabularnewline
    \hline
    \hline
     0-1 & 0.6974 & 0.6767 & 0.5399 & 0.5207 \\
     0-2 & 0.1354 & 0.1407 & 0.1634 & 0.1393 \\
     0-3 & 0.1283 & 0.1283 & 0.1306 & 0.1112 \\
     1-2 & 0.4536 & 0.4329 & 0.3880 & 0.3053 \\
     1-3 & 0.3258 & 0.3253 & 0.1574 & 0.1113 \\
     2-3 & 0.7216 & 0.6812 & 0.6210 & 0.4986 \\
    \hline
    \hline
    \multicolumn{5}{r}{Overlap Reduction Relative to Y3 SOM \textit{riz}}\tabularnewline
    \hline
    \hline
     0-1 & - & 3$\%$  & 23$\%$  & 25$\%$ \\
     0-2 & - & -4$\%$ & -21$\%$ & -3$\%$ \\
     0-3 & - & 0$\%$  & -2$\%$  & 13$\%$ \\
     1-2 & - & 5$\%$  & 14$\%$  & 33$\%$ \\
     1-3 & - & 0$\%$  & 52$\%$ & 66$\%$ \\
     2-3 & - & 6$\%$  & 14$\%$ & 31$\%$ \\
    \hline
    \hline
    \end{tabular}
    \caption{Amount of bin overlap between each redshift bin pair, for each method, together if the percentage overlap reduction with respect to the Y3 SOM \textit{riz} (the fiducial method used in DES Y3).}
    \label{tab:overlap}
\end{table}

\section{Impact on cosmological parameters}\label{sec:results:cosmo}

In this Section we quantify how the changes in the SOMPZ method proposed in this paper impact the final cosmological constraints. In particular, we want to see how these changes impact the $S_8-\omegam$ plane, i.e., the main cosmology results for DES Y3.

Changing the redshift estimation of the source catalog impacts all the following steps in the cosmology estimation pipeline. 
In the SOMPZ method (see Section \ref{sec:som}), the galaxy bin assignment is based on the wide SOM assignment. 
Therefore, when we train a new wide SOM, the galaxies are re-assigned and it is necessary to re-compute the two-point statistics measurements and the covariance matrix. 
In the case of the SOMF method with the \textit{riz} bands, it is possible to perform all those steps using the Y3 data, and get a direct comparison between the Y3 SOM and the SOMF in the Y3 cosmology. 
In the cases when we add the \textit{g}-band information, both for the Y3 SOM and the SOMF, it is not possible to carry the comparison all the way to the cosmological parameters, given that we do not have shape measurements for the \textit{g}-band. Instead, we generated simulated datavectors, based on the Y3 cosmology, and compare the contours obtained in this simulated data. 

Since in this paper we are exploring modifications on the method for redshift estimation for the weak lensing source catalog, we will be focusing on the cosmic shear measurement. Notice however, that the changes discussed here also impact galaxy-galaxy lensing and, naturally, the $3\times2$pt statistics.

\subsection{Cosmological Constraints - Y3 Data}\label{sec:results:cosmo_intro}

We tested the impact of replacing the SOM algorithm all the way from the SOM creation and assignment, to the cosmological parameter estimation. We use the redshift estimation schema and data described in Section \ref{sec:sompz}, but replace the SOM algorithm outline in Section \ref{sec:fiducial} (see also \citealt*{DES:2019bxr}), with the one outlined in Section \ref{sec:somf} (see also \citealt*{Sanchez2020}). 

Creating a new wide SOM and assigning the wide sample to it has a significant impact on bin assignments, influencing which galaxies are sorted into specific redshift bins. Consequently, to derive cosmological parameters, it becomes necessary to recalibrate various components of the analysis. This entails the reassessment of 2-point statistics, and the subsequent re-calculation of the covariance matrix. In essence, this process entails a complete reconstruction of the data-vector. Details on each step can be found in Appendix \ref{sec:shear_meas}.
Subsequently, we initiate a parameter estimation chain using the updated data-vector, adhering to the methodology outlined in the DES Y3 pipeline, as comprehensively expounded in \citet*{krause21}, and concisely summarized in Appendix \ref{sec:theory}. It's worth emphasizing that we follow the same methodology as \citet*{DES:2021bvc} and \citet*{DES:2021vln}.

\begin{figure}
\centering
\includegraphics[width=\linewidth]{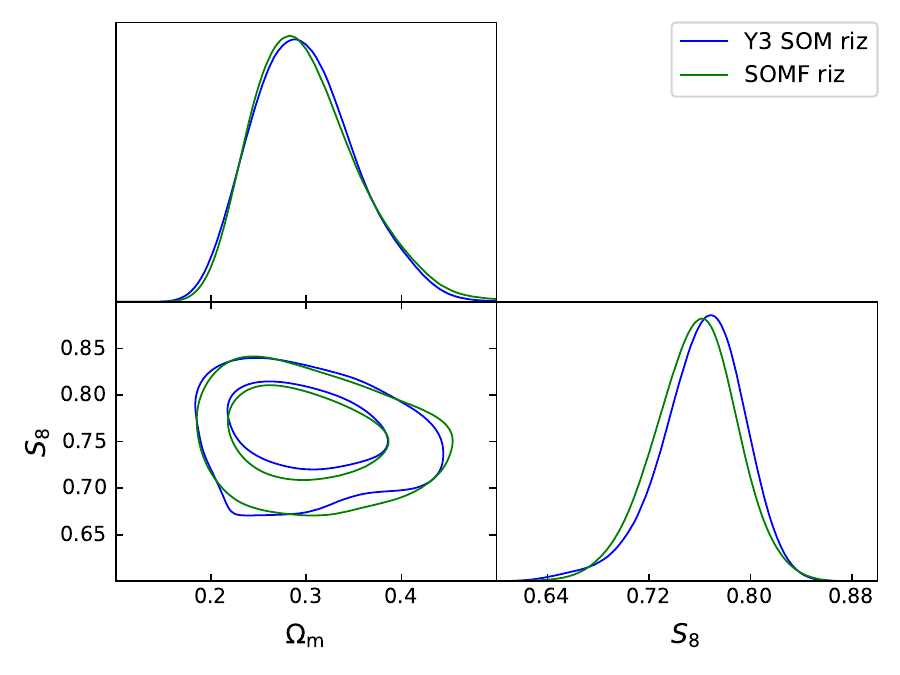}
\caption{Cosmological constraints on the clustering amplitude, $S_8$ with the matter density, $\Omega_m$ in $\Lambda$CDM, using the DES Y3 data. The marginalised posterior contours (inner $68\%$ and outer $95\%$ confidence levels) are shown for the Y3 SOM in blue and SOMF in green. 
}
\label{fig:somf_cosmo_oms8}
\end{figure}

Figure \ref{fig:somf_cosmo_oms8} compares the $1\sigma$ and $2\sigma$ contours in the $\Omega_m - \sigma_8$ plane. The blue contour uses the Y3 SOM algorithm, however, containing only the SOMPZ information when constraining the redshift (as opposed to the complete redshift information used in Y3 that contains SOMPZ + Clustering Redshifts (WZ) + Blending + Shear Ratios information, as detailed in \citealt*{DES:2020ebm}). 
The green contour shows the chain for which the data vector was constructed with the redshift information from the SOMF algorithm. 

The contours agree at the level of chain variance, and we do not observe any gain in constraining power on the cosmological parameters, due to the small reduction in bin overlap obtained when using the SOMF \textit{riz} (see Figure \ref{fig:overlap}). The marginalized mean $S_8$ and $\omegam$ values in \lcdm \ are:

\begin{eqnarray*}
S_{8} & = & 0.761_{-0.027}^{+0.037}\,\quad \quad\quad (\textrm{Y3 SOM})\\
\Omega_{\rm m} & = & 0.298_{-0.061}^{+0.046}\,\quad \quad\quad (\textrm{Y3 SOM})
\end{eqnarray*}
and
\begin{eqnarray*}
S_{8} & = & 0.756_{-0.030}^{+0.035}\,\quad
\quad\quad   (\textrm{SOMF}) ~~~\\
\Omega_{\rm m} & = & 0.301_{-0.066}^{+0.041}\,\quad \quad\quad   (\textrm{SOMF}) ~~~
\end{eqnarray*}

\noindent where uncertainties are 68\% confidence intervals. We can see that the values agree well within the uncertainty level, and the confidence intervals are also equivalent. Notice that the mean $S_8$ and $\omegam$ for the Y3 SOM are not the same as the ones quoted in \cite*{DES:2021bvc} and \cite*{DES:2021vln}, given that here our chains includes redshifts estimated only with the SOMPZ method, but again they are in perfect agreement. 

The good level of agreement of the two chains, and to the Y3 fiducial results, demonstrates the robustness of our method, one of the main results of this paper. This result, combined to the agreement in mean redshift and shape of the distribution, demonstrates that the SOMF algorithm is compatible with the SOMPZ pipeline, and robust against the cosmology results, validating it and making it a viable option for DES Year 6. 

We also emphasize that the improvements on cosmology due to the enhanced redshift methodology described in this paper could be more significant for a cosmic shear analysis more limited by redshift uncertainty than DES Y3.

\subsection{Cosmological Constraints - Simulations}

To assess the potential impact of incorporating \textit{g}-band information into redshift estimation on cosmological parameters, we constructed simulated data-vectors based on the DES Y3 setup, as detailed in Appendix \ref{sec:data:gen_data}. Subsequently, we analyzed these simulations using the Y3 pipeline.
We generated simulated data for four distinct cases, each employing different methods: the Y3 SOM method with the \textit{riz} bands, the SOMF method with the \textit{riz} bands, the Y3 SOM method with the \textit{griz} bands, and the SOMF method with the \textit{griz} bands. 

At the level of simulated data vectors we can already see differences. Switching from using the \textit{riz} SOM $n(z)$s to \textit{griz} SOMF, we see a rougly 5-10$\%$ increase in the lensing ($\kappa \kappa$) signal in the uppermost $4,4$ bin correlation. This is likely due to the small upwards shift in the mean, and the reduction in the weight of the low redshift tail. Much of the signal-to-noise of cosmic shear comes from these upper bin correlations, and so boosting the signal here is useful for optimising our cosmological constraint.
We also see an overall reduction of the intrinsic alignment II contribution in the cross-bin correlations, as well as a decrease of the GI component in the auto bin pairs. This is again expected from a reduction in the width of the source bins (we can see this by considering, for example, Eq 16-17 of \citealt*{DES:2021vln}). The cleaner separation of IA signals is helpful as it can break degeneracies and allow the data to constrain IAs more effectively \citep{campos22}.

\begin{figure}
\centering
\includegraphics[width=\linewidth]{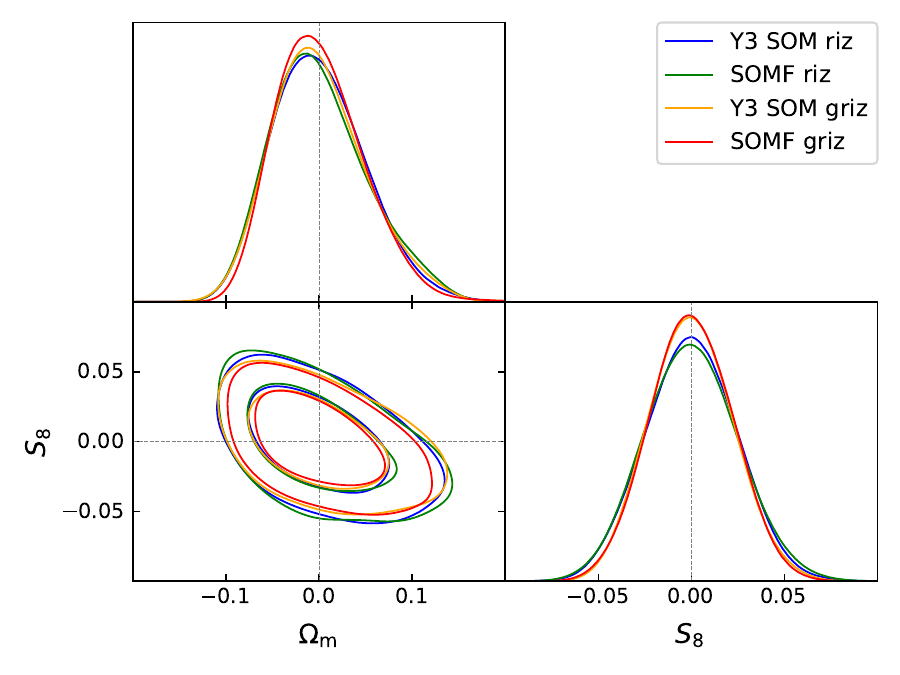}
\caption{Cosmological constraints on the clustering amplitude, $S_8$ with the matter density, $\Omega_m$ in $\Lambda$CDM, using the simulated data described in Section \ref{sec:data:gen_data}, in order to include the \textit{g}-band information. The marginalised posterior contours (inner $68\%$ and outer $95\%$ confidence levels) are shown for the Y3 SOM in blue and SOMF in green, for the \textit{riz} bands, and Y3 SOM in yellow and SOMF in red for the  \textit{griz} bands.
Notice that the parameters have been centered to zero, in order to focus on the relative error reduction.}
\label{fig:sim_oms8}
\end{figure}
\begingroup

\setlength{\tabcolsep}{10pt} 
\renewcommand{\arraystretch}{1.5} 

\begin{table}
    \centering
    \caption{Values of the 68$\%$ confidence intervals, shown in Figure \ref{fig:sim_oms8}, for the Y3 SOM and SOMF using \textit{riz} bands, and Y3 SOM and SOMF using \textit{griz} bands.}

    \begin{tabular}{l|c|c}
          & $\Delta S_{8}$ & $\Delta\Omega_{\rm m}$ \\
         \cline{1-3}
         Y3 SOM \textit{riz} & $\pm 0.024$ & $_{-0.056}^{+0.044}$ \\
         \cline{1-3}
         SOMF \textit{riz} & $\pm 0.025$ & $_{-0.061}^{+0.041}$  \\
         \cline{1-3}
         Y3 SOM \textit{griz} & $\pm 0.022$ & $_{-0.058}^{+0.042}$  \\
        \cline{1-3}
         SOMF \textit{griz} & $\pm 0.022$ & $_{-0.054}^{+0.039}$  \\

    \end{tabular}
    \label{tab:sim_delta}
\end{table}

\endgroup

In Figure \ref{fig:sim_oms8}, we present a comparison of the $1\sigma$ and $2\sigma$ contours in the $\Omega_m - \sigma_8$ plane for each case. Importantly, the cosmological parameters have been centered to zero in these plots, enabling us to concentrate on the gain in signal-to-noise (SN) or the reduction in errors relative to one another. This centered approach helps highlight how the various methods improve in comparison to their counterparts. Table \ref{tab:sim_delta} shows the numerical values of the 1$\sigma$ uncertainty intervals for each case shown in Figure \ref{fig:sim_oms8}.

\begin{figure}
\centering
\includegraphics[width=\linewidth]{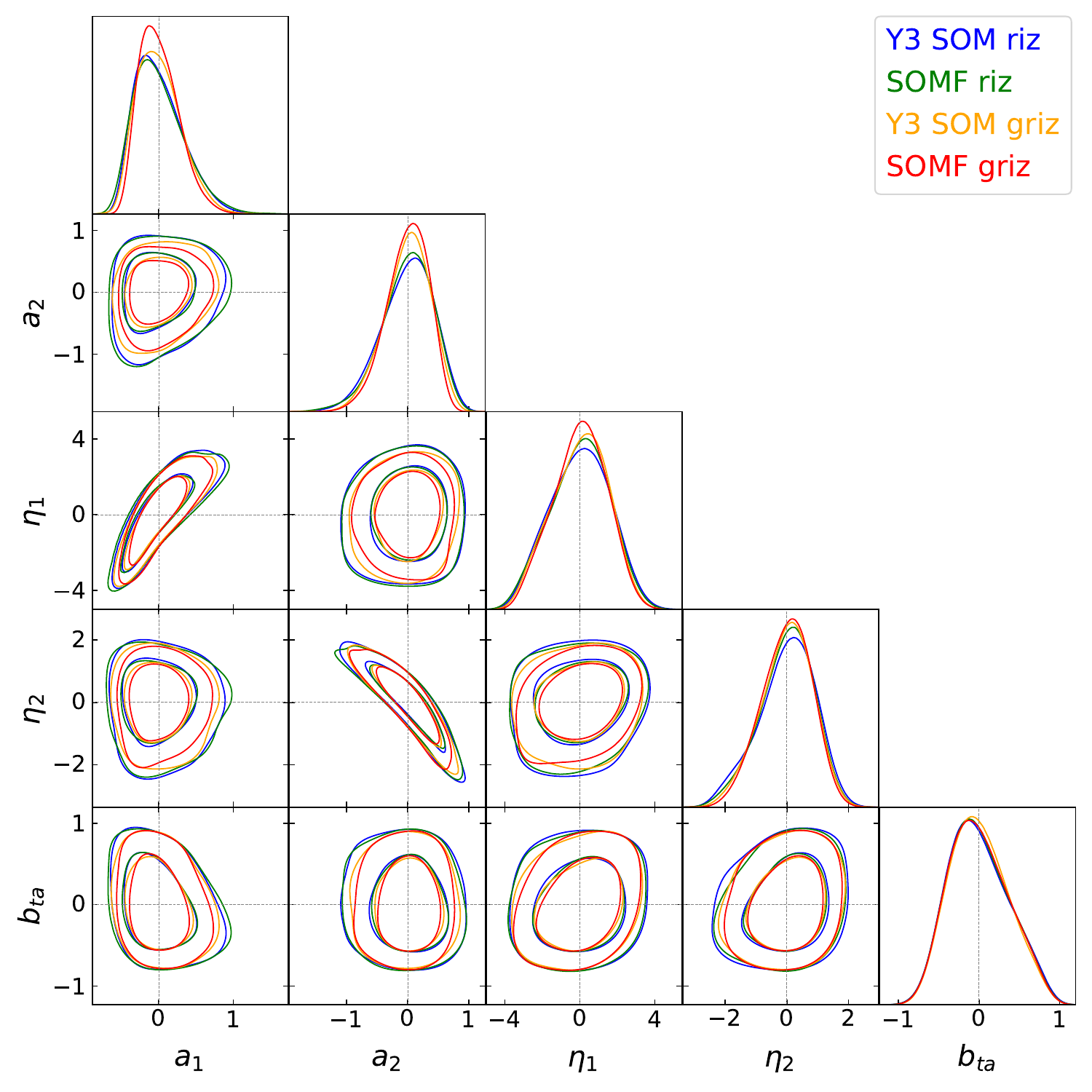}
\caption{Intrinsic Alignment parameters constraints in $\Lambda$CDM, using the simulated data described in Section \ref{sec:data:gen_data}, in order to include the \textit{g}-band information. The marginalised posterior contours (inner $68\%$ and outer $95\%$ confidence levels) are shown for the Y3 SOM in blue and SOMF in green, for the \textit{riz} bands, and Y3 SOM in yellow and SOMF in red for the  \textit{griz} bands.
Notice that the parameters have been centered to zero, in order to focus on the relative error reduction.}
\label{fig:sim_ia}
\end{figure}

Upon examination of Figure \ref{fig:sim_oms8} and Table \ref{tab:sim_delta}, we conclude that the SOMF method with \textit{griz} bands exhibits the most significant improvement in constraining both $\Omega_m$ and $S_8$, reducing the error bars by $\sim 10\%$, in each parameter, over Y3 SOM \textit{riz}. Moreover, even though both the Y3 SOM and SOMF methods using \textit{griz} bands demonstrate comparable constraining power for $S_8$, the SOMF shows better constraints in $\Omega_m$, highlighting the effectiveness of the SOMF algorithm. 
Additionally, Figure \ref{fig:sim_ia} presents the constraints on Intrinsic Alignment parameters, further underscoring the benefits of adding \textit{g}-band data and using the SOMF algorithm, as we expected from the datavector. 

Consequently, we recommend incorporating the SOMF \textit{griz} method in the upcoming DES Year 6 analysis. While the primary gain in constraining power currently stems from the addition of \textit{g}-band data, this trend might evolve in future surveys as we observe more faint objects.

\section{Conclusions}

In this paper, we explore three modifications to the SOMPZ method (\citealt*{DES:2019bxr}, \citealt*{DES:2020ebm}) employed in the DES Year 3 analysis: 1) changing the SOM algorithm; 2) including the photometry from the \textit{g}-band; 3) adding the redshift information, when available, to train and assign galaxies to the SOM.
Our goal is to optimize our redshift estimation pipeline for the DES Year 6 data, with especial focus on the weak lensing source galaxies. This is an important problem, given that those galaxies will be deeper and fainter compared to DES Y3, and we want to be able to treat them properly, minimizing cuts in our catalogs.
That said, the findings of this paper are applicable to ensemble redshift estimation in general, and can be used as a guide for the lens redshifts in DES Year 6, and the redshift analysis of other surveys. 

Using the DES Y3 weak lensing data, we tested each of the three modifications, and compared their impact relative to the Year 3 fiducial results.
The main conclusions of our study are as follows:

\begin{itemize}

\item We showed that the SOMF successfully compresses the high-dimensional flux space into the redshift space in a smooth way, i.e., transitions between low- and high-redshift happen gradually. Given that Self-Organizing Map is a unsupervised clustering algorithm, neighbor cells should present similar properties. This property should definitely be observed in the flux space (the features the SOM was trained on, Figures \ref{fig:wide_somf_y3} and \ref{fig:deep_somf_y3} ), and the fact that its also present in redshift space, Figure \ref{fig:somf_deep_wide_std}, is evidence of the successful mapping of redshifts. 

\item By using the SOMF algorithm, tailored for the problem of photometric redshifts, we were able to reduce the standard deviation in the redshift distribution in each cell. Figure \ref{fig:std_somf_y3som} shows a $\sim 20\%$ reduction in the median of the standard deviation for SOMF when compared to the Y3 SOM. That means that, within each cell of the SOM, we have a better estimation of the redshift of those galaxies.  

\item We are able to reduce bin-overlap even further, and therefore have better defined ensemble redshift distributions. Using the SOMF algorithm helps, however including an additional photometric band, the \textit{g}-band, plays a major role in reducing the overlap between redshift bins.  

\item Two-point measurements and cosmology results are robust. Changing the galaxy assignment in the wide SOM affects all the subsequent stages of the pipeline, meaning that all measurements need to be repeated. We tested the robustness of the 2pt cosmic shear measurements, and the final cosmology results, verifying that they are completely consistent with our Y3 findings. Any observed shift of $\Omega_m$ and $S_8$ was within previous systematic uncertainty. In addition, SOMF \textit{griz} reduces the error bars in each of these parameters by $\sim 10\%$ over Y3 SOM \textit{riz}.

\item We considered adding the redshift measurements that we had available for galaxies in our spectroscopic sample as an additional feature to train the SOM and assign galaxies to it. This path did not lead to improvements in the method, as shown in Appendix \ref{sec:appendixA}, since the inclusion of the redshift feature seem to dominate over the other features, and create a very sparse SOM. 

\end{itemize}

\noindent

Photometric redshift estimation is an important topic in cosmology. There have been several proposals on how to improve redshift estimation, given the limitations imposed by photometric data. Self-Organizing Maps, in particular the SOMPZ pipeline proposed for DES Year 3 showed promising results, constraining the redshift of the weak lensing source galaxies to the $2\%$ level. That was a ground-breaking work, that leverages on the deep fields high-quality photometric data to connect the spectroscopic information available for a small group of galaxies, to the main wide data set.

This paper sets out an recommendation for improvement of the redshift estimation pipeline, SOMPZ, for weak leasing source galaxies in DES Year 6.
Although the fiducial SOMPZ method employed for Y3 is perfectly suitable for Y6 as well, by switching the Y3 SOM recipe to the SOMF and including the \textit{g}-band photometry (even if only at the redshifts level) we can obtain even better redshift estimates for the Y3 wide sample. We expect these effects to be even more accentuated for Y6, given the increased depth of the wide sample, therefore the changes proposed here will have an even greater impact.

The implications of our results extend beyond the DES Y6 project. They provide a valuable foundation for the improvement and refinement of redshift characterization in future Stage IV surveys, such as those conducted by the Rubin Observatory, Euclid Collaboration and Roman Space Telescope. By reducing redshift bin overlap and enhancing the accuracy of redshift estimates, we are poised to unlock new possibilities for advancing our understanding of the universe's dark components, and to achieve more precise and robust cosmological parameter estimates in the years to come.

It is also worth emphasising that the code for the SOMPZ pipeline including the SOMF algorithm was further revised, simplified and documented, making it easier to use. Given its generality in the redshift estimation context, simplicity, and open source nature, we foresee the use of this method in future analyses, as a ensemble photometric redshift estimation pipeline option for many data sets.

\section*{Code and Data Availability}

The photometric redshift ensemble estimation code used in this work is publicly available at \url{https://github.com/AndresaCampos/sompz_y6}.

The DES Y3 data products used in this work, are publicly available at https://des.ncsa.illinois.edu/releases. As cosmology likelihood sampling software we use \texttt{cosmosis}, available at https://github.com/joezuntz/cosmosis.

\section*{Contribution Statement}

AC: Designed the project, developed the code, led the analysis and the writing of the manuscript. BY: Performed the analysis of SOMz, merged \textit{g}-band data and edited the manuscript. SD: Supervised the project, assisted with result interpretation and edited the manuscript. AA: Provided suggestions and feedback on analyses, assisted with result interpretation and edited the manuscript. AA: Assisted with project design, code testing and review, and result interpretation. CS: Assisted with project design and result interpretation. GB: Proposed the project, provided updated SOM algorithm, assisted with result interpretation.  
GG, JM, SS: DES internal reviewers of the paper. 
Production of the DES gold catalog: ISN, KB, MCK, ACR, MB, ADW, RG, ER, ES, BY. 
Production of the DES shape catalog: MG, ES, AA, MB, MT, AC, CD, NM, ANA, IH, DG, GB, MJ, LFS, AF, JM, RPR, RC, CC, SP, IT, JP, JEP, CS. 
Production of the DES PSF: MJ, GB, AA, CD, PFL, KB, IH, MG, AR. 
Production of the DES deep fields: WH, AC, AA, RG, ES, IH, GB, ISN, BY, KE, TD. 
Production of the DES Balrog: SE,BY, NK, EH, YZ. 
Production of the DES redshift distribution: JM, AA, AA, CS, SE, JD, JM, DG, GB, MT, SD, AC, NM, BY, MR. 
Production of the DES image simulations: NM, MB, JM, AA, DG, MJ, AC, MT, ES, BY, KH, SD, JZ, KE, RR, TV.
Production of the DES covariance: OF, FAO, HC, OA, RR, JS, XF, TE, EK.
Production of the DES methods: EK, XF, SP, LS, OA, HH, JB, JP, JZ, TE, JD.
Production of the sims: JD, RW, RB, ER, SP, NM, AA, JM.
Builders: The remaining authors have made contributions to this paper that include, but are not limited to, the construction of DECam and other aspects of collecting the data; data processing and calibration; developing broadly used methods, codes, and simulations; running the pipelines and validation tests; and promoting the science analysis.

\section*{Acknowledgements}

Andresa Campos thanks the support from the U.S. Department of Energy grant DE-SC0010118 and the NSF AI Institute: Physics of the Future, NSF PHY-2020295. 

Funding for the DES Projects has been provided by the U.S. Department of Energy, the U.S. National Science Foundation, the Ministry of Science and Education of Spain, 
the Science and Technology Facilities Council of the United Kingdom, the Higher Education Funding Council for England, the National Center for Supercomputing 
Applications at the University of Illinois at Urbana-Champaign, the Kavli Institute of Cosmological Physics at the University of Chicago, 
the Center for Cosmology and Astro-Particle Physics at the Ohio State University,
the Mitchell Institute for Fundamental Physics and Astronomy at Texas A\&M University, Financiadora de Estudos e Projetos, 
Funda{\c c}{\~a}o Carlos Chagas Filho de Amparo {\`a} Pesquisa do Estado do Rio de Janeiro, Conselho Nacional de Desenvolvimento Cient{\'i}fico e Tecnol{\'o}gico and 
the Minist{\'e}rio da Ci{\^e}ncia, Tecnologia e Inova{\c c}{\~a}o, the Deutsche Forschungsgemeinschaft and the Collaborating Institutions in the Dark Energy Survey.

The Collaborating Institutions are Argonne National Laboratory, the University of California at Santa Cruz, the University of Cambridge, Centro de Investigaciones Energ{\'e}ticas, 
Medioambientales y Tecnol{\'o}gicas-Madrid, the University of Chicago, University College London, the DES-Brazil Consortium, the University of Edinburgh, 
the Eidgen{\"o}ssische Technische Hochschule (ETH) Z{\"u}rich, 
Fermi National Accelerator Laboratory, the University of Illinois at Urbana-Champaign, the Institut de Ci{\`e}ncies de l'Espai (IEEC/CSIC), 
the Institut de F{\'i}sica d'Altes Energies, Lawrence Berkeley National Laboratory, the Ludwig-Maximilians Universit{\"a}t M{\"u}nchen and the associated Excellence Cluster Universe, 
the University of Michigan, the National Optical Astronomy Observatory, the University of Nottingham, The Ohio State University, the University of Pennsylvania, the University of Portsmouth, 
SLAC National Accelerator Laboratory, Stanford University, the University of Sussex, Texas A\&M University, and the OzDES Membership Consortium.

Based in part on observations at Cerro Tololo Inter-American Observatory at NSF's NOIRLab (NOIRLab Prop. ID 2012B-0001; PI: J. Frieman), which is managed by the Association of Universities for Research in Astronomy (AURA) under a cooperative agreement with the National Science Foundation.

The DES data management system is supported by the National Science Foundation under Grant Numbers AST-1138766 and AST-1536171.
The DES participants from Spanish institutions are partially supported by MINECO under grants AYA2015-71825, ESP2015-66861, FPA2015-68048, SEV-2016-0588, SEV-2016-0597, and MDM-2015-0509, 
some of which include ERDF funds from the European Union. IFAE is partially funded by the CERCA program of the Generalitat de Catalunya.
Research leading to these results has received funding from the European Research
Council under the European Union's Seventh Framework Program (FP7/2007-2013) including ERC grant agreements 240672, 291329, and 306478.
We  acknowledge support from the Brazilian Instituto Nacional de Ci\^encia
e Tecnologia (INCT) e-Universe (CNPq grant 465376/2014-2).

This manuscript has been authored by Fermi Research Alliance, LLC under Contract No. DE-AC02-07CH11359 with the U.S. Department of Energy, Office of Science, Office of High Energy Physics.

\bibliography{refs.bib}
\bibliographystyle{mnras_2author}

\newpage

\appendix
\section{Magnitude and Colors - SOMF}

\begin{figure*}
\centering
\includegraphics[width=2.\columnwidth]{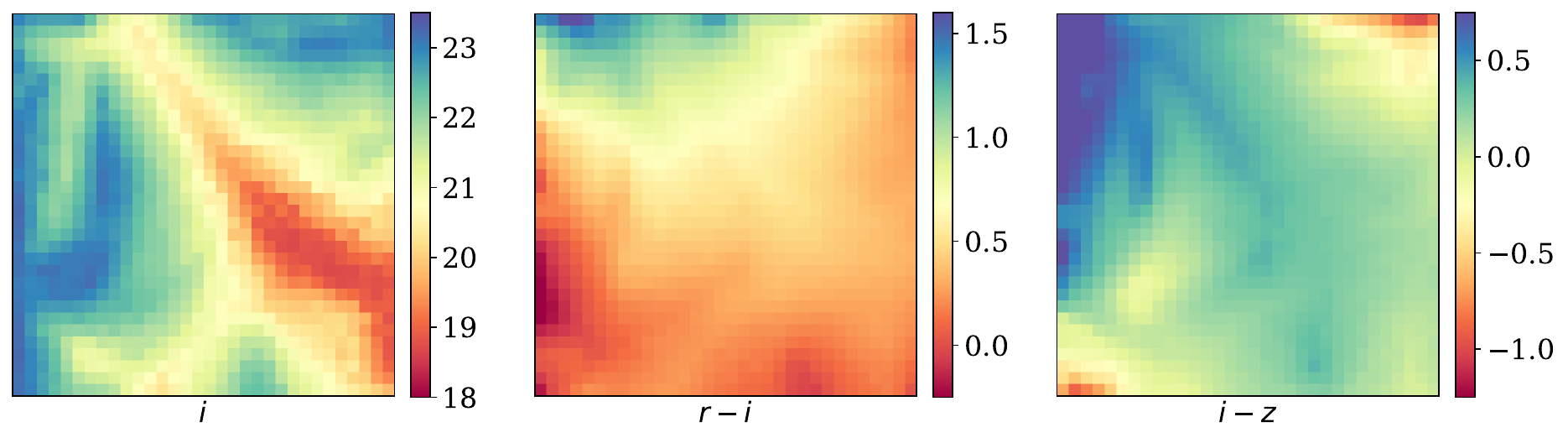}
\caption{Wide Self-Organizing Map constructed using the SOMF algorithm and data from the \textit{riz} bands. The visualization depicts the mean $i$-band magnitude (on the left), the mean $r$ - $i$ color (in the middle), and the mean $i$ - $z$ color (on the right) for each cell within the wide SOM.}
\label{fig:wide_somf_y3}
\end{figure*}

\begin{figure*}
\centering
\includegraphics[width=2.\columnwidth]{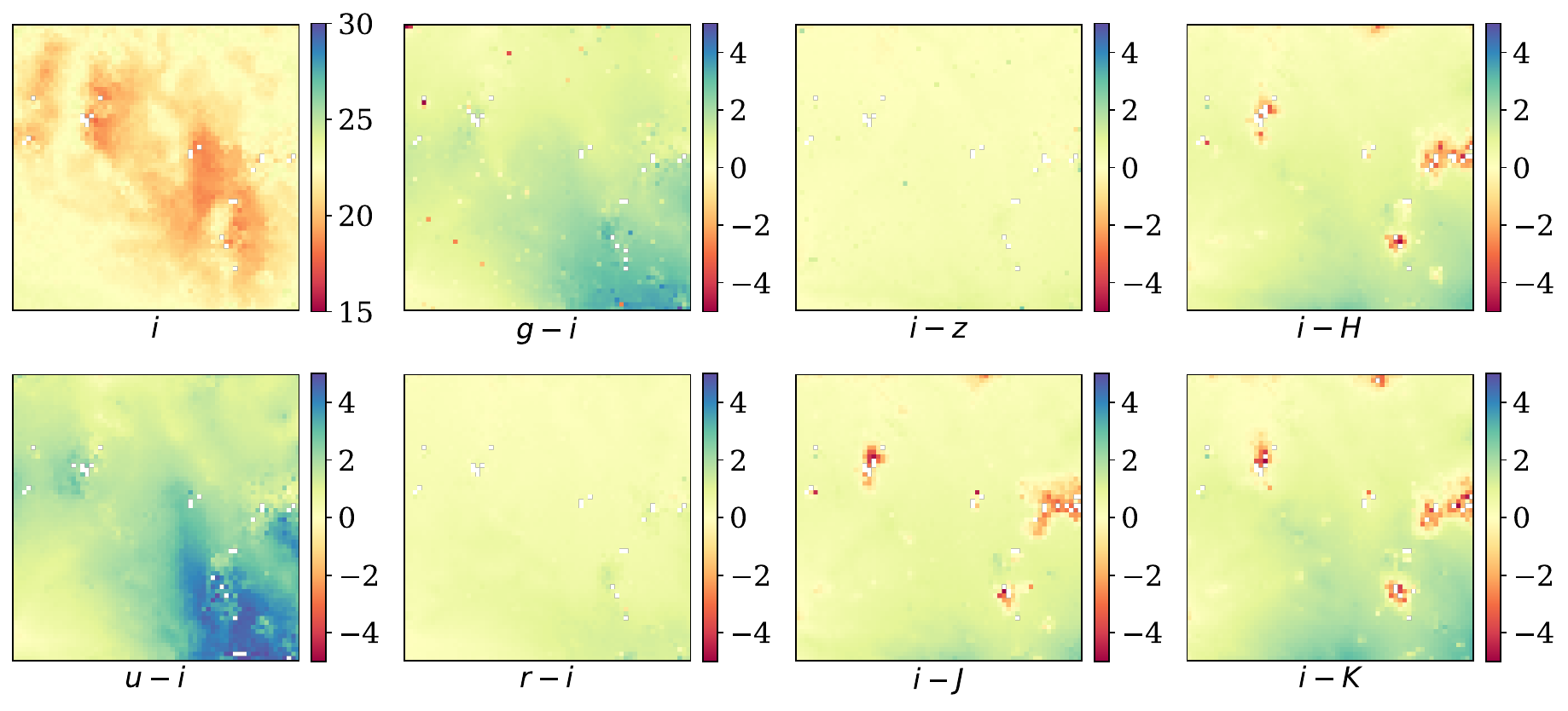}
\caption{Deep field Self-Organizing Map constructed using the SOMF algorithm with data from the $ugrizJHK$ bands. In the upper left, we have the mean $i$-band magnitude for each cell within the deep SOM. Additionally, the various colors utilized in the deep SOM training are shown.}
\label{fig:deep_somf_y3}
\end{figure*}

Figures \ref{fig:wide_somf_y3} and \ref{fig:deep_somf_y3} provide visual representations of the \textit{i}-band magnitude and colors for each cell within the wide and deep Self-Organizing Map (SOM), respectively. These maps were created using the SOMF algorithm, as  outlined in Section \ref{sec:somf}. 
The SOM is designed to create a smooth map encompassing the entire parameter space derived from the training inputs. By drawing a comparison with Figure \ref{fig:somf_deep_wide_std}, we can appreciate how effectively the SOMF method maps the color-redshift relationship. This effectiveness is evident through the creation of a smooth redshift map that corresponds with the observed color patterns. The correlation between color and redshift in the SOM effectively illustrates its capacity to capture and represent this intricate relationship.
Any abrupt differences observed between adjacent cells can be interpreted as indirect indications of potential degeneracies within the color-redshift relationship.

\section{SOM-z}
\label{sec:appendixA}

\begin{figure*}
    \centering
    \includegraphics[width=0.86\linewidth]{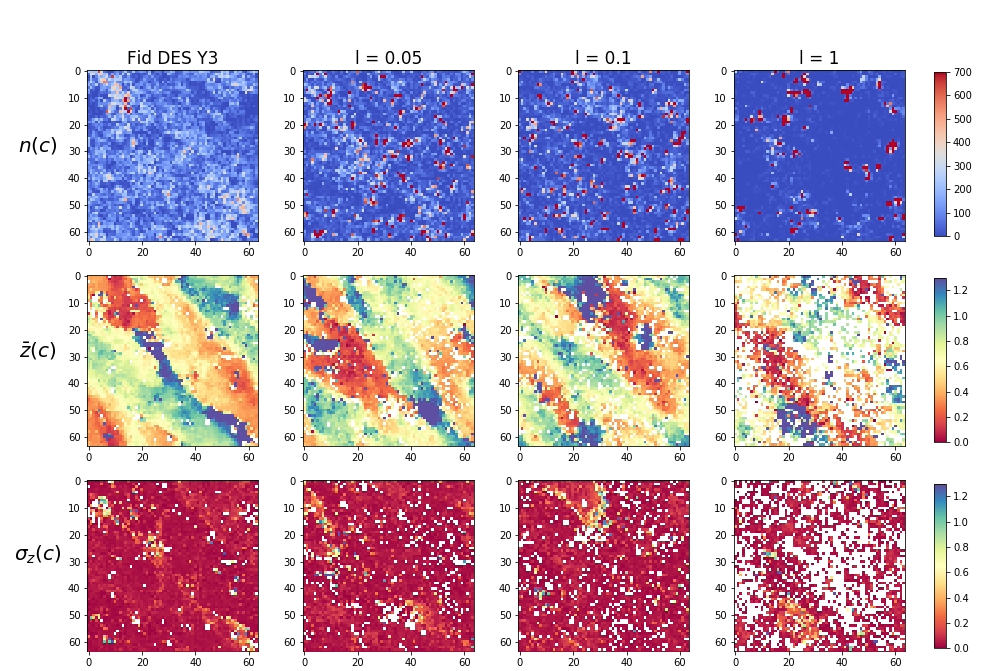}
    \caption{Adding redshift information in training and assigning deep SOM, using Y3 SOM.}
    \label{fig:som addz}   
\end{figure*}

\begin{figure*}
    \centering
    \includegraphics[width=0.86\linewidth]{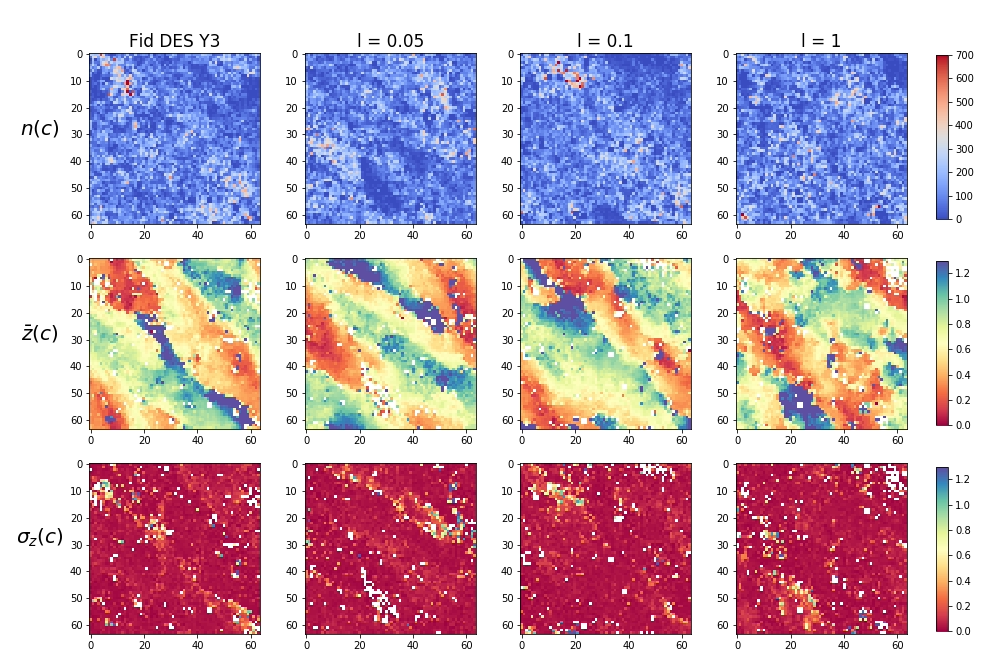}
    \caption{Adding redshift information in training deep SOM only, using Y3 SOM.}
    \label{fig:som tzaf}
\end{figure*}

In this section, we delve into the details of incorporating redshift information, when available, as an extra feature in the training and assignment of the deep SOM used in the SOMPZ method. This is an unconventional way of applying an unsupervised learning method, such as the SOM, however we were motivated to investigate whether this approach would maximize the use of the available information, leading to enhancements in our ability to estimate redshifts. We applied this strategy to the Y3 SOM, described in Section \ref{sec:som}. In light of what we observed, described in what follows, we did not try applying this strategy to the SOMF. 

An interesting phenomenon arises when we include redshift information in the training and assignment process. The redshift sample galaxies tend to cluster in a few cells, leaving most cells sparsely occupied. This behavior can be observed in Figure~\ref{fig:som addz} (top), where we show the number of spectroscopic galaxies assigned to each cell in the deep SOM (top row), the mean redshift per cell (middle row), and the standard deviation in each cell (bottom row). The first column shows the DES Y3 SOM, without including the redshift information, and $\lambda$ is a scaling factor for the contribution of redshift information relative to flux data, i.e., $\lambda = 0$ means no redshift contribution, while $\lambda=1$ means the contribution of the redshift is the same as a flux. As the contribution of redshift information increases, the clustering effect intensifies. 

\begin{figure}
\centering
\includegraphics[width=\linewidth]{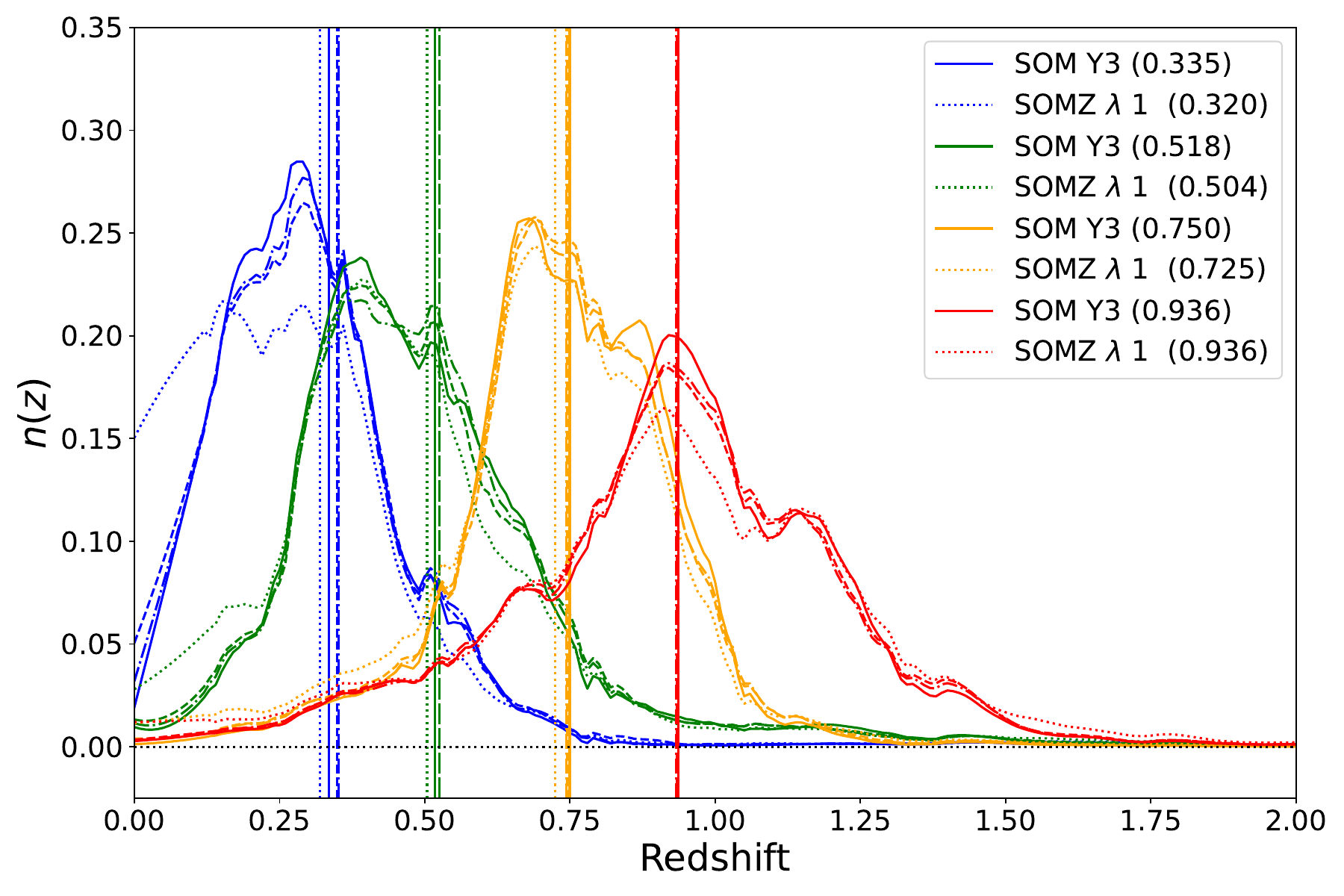}
\caption{Comparison of the \textit{n(z)} bins obtained using the fiducial DES Y3 SOM (solid line), and adding redshift in training and assigning deep SOM. The dotted line represents the most "extreme" case, where $\lambda = 1$ and the contribution of the redshift in training and assigning is the same as the fluxes, while the dashed and dot-dashed lines represent $\lambda = 0.1$ and $\lambda = 0.05$ respectively. The vertical lines are the mean redshift in each bin, shown in the legend for the fiducial method, or $\lambda = 0$,  and the $\lambda=1$ case.}
\label{fig:nz add z}
\end{figure}

\begin{figure}
\centering
\includegraphics[width=\linewidth]{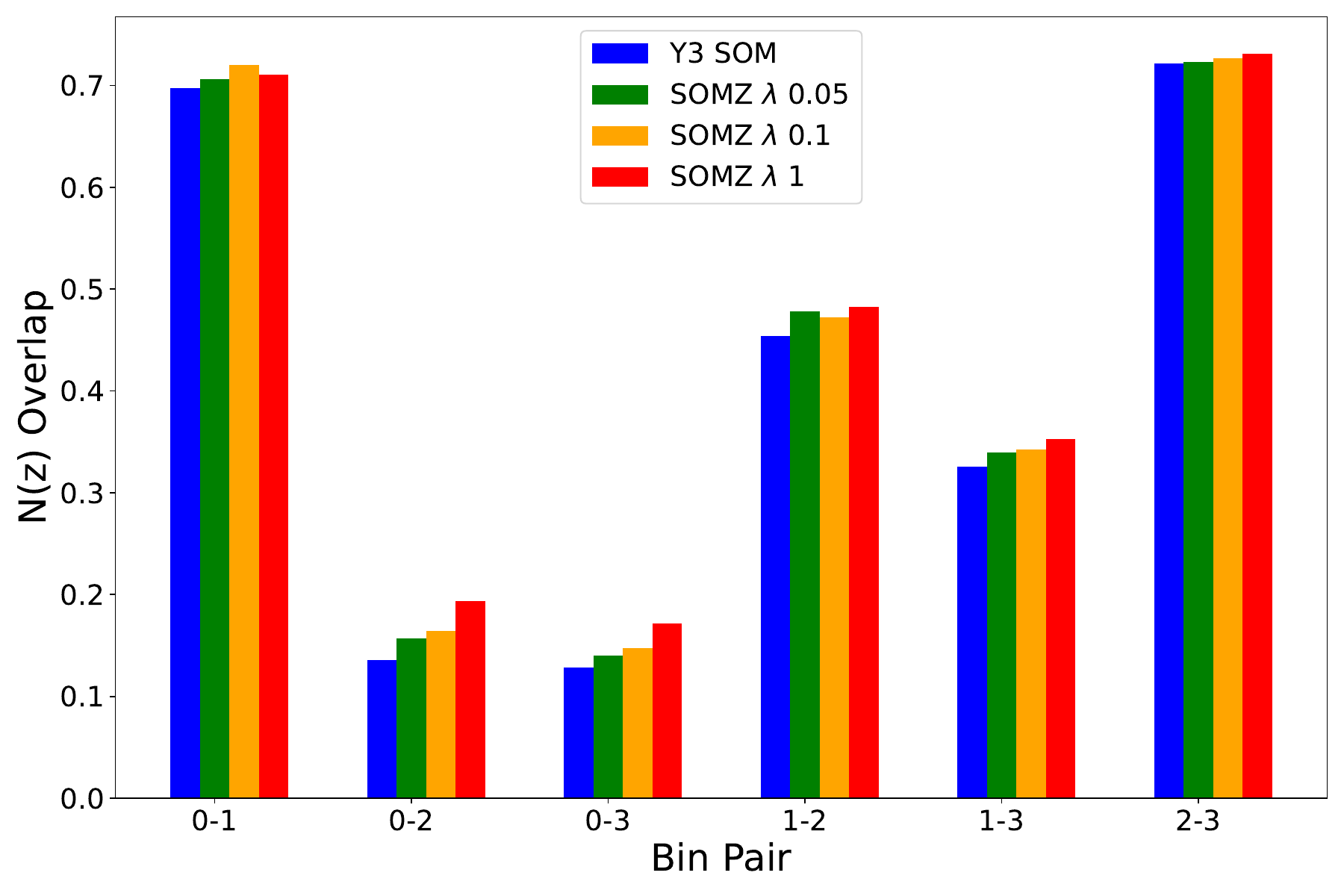}
\caption{Redshfit bin overlap plot for fiducial DES Y3 (blue) and adding redshift in both training and assigning deep SOM. The bin overlap increases as the contribution of the redshift, represented by $\lambda$, increases. The green line represents $\lambda = 0.05$ or $5\%$ contribution, the yellow $\lambda = 0.1$, contributing $10\%$, and the red line $\lambda=1$, contributing the same as flux.}
\label{fig:overlap add z}
\end{figure}

This clustering of galaxies on the deep SOM negatively impacts the photometric redshift calibration. Figure~\ref{fig:nz add z} presents the wide data photometric distribution for the DES Y3 method and the variants with added redshift information. 
The solid line represents the DES Y3 $n(z)$, while the 
dot-dashed and dashed lines represent $\lambda = 0.05$ and $\lambda = 0.1$ respectively, and the dotted line represents the most "extreme" case, where $\lambda = 1$. As redshift information gains more weight, the $n(z)$ distribution in each redshift bin spreads further. This results in increased bin overlap, as illustrated in Figure~\ref{fig:overlap add z}, which is the opposite effect of what we are looking for.

\begin{figure}
\centering
\includegraphics[width=\linewidth]{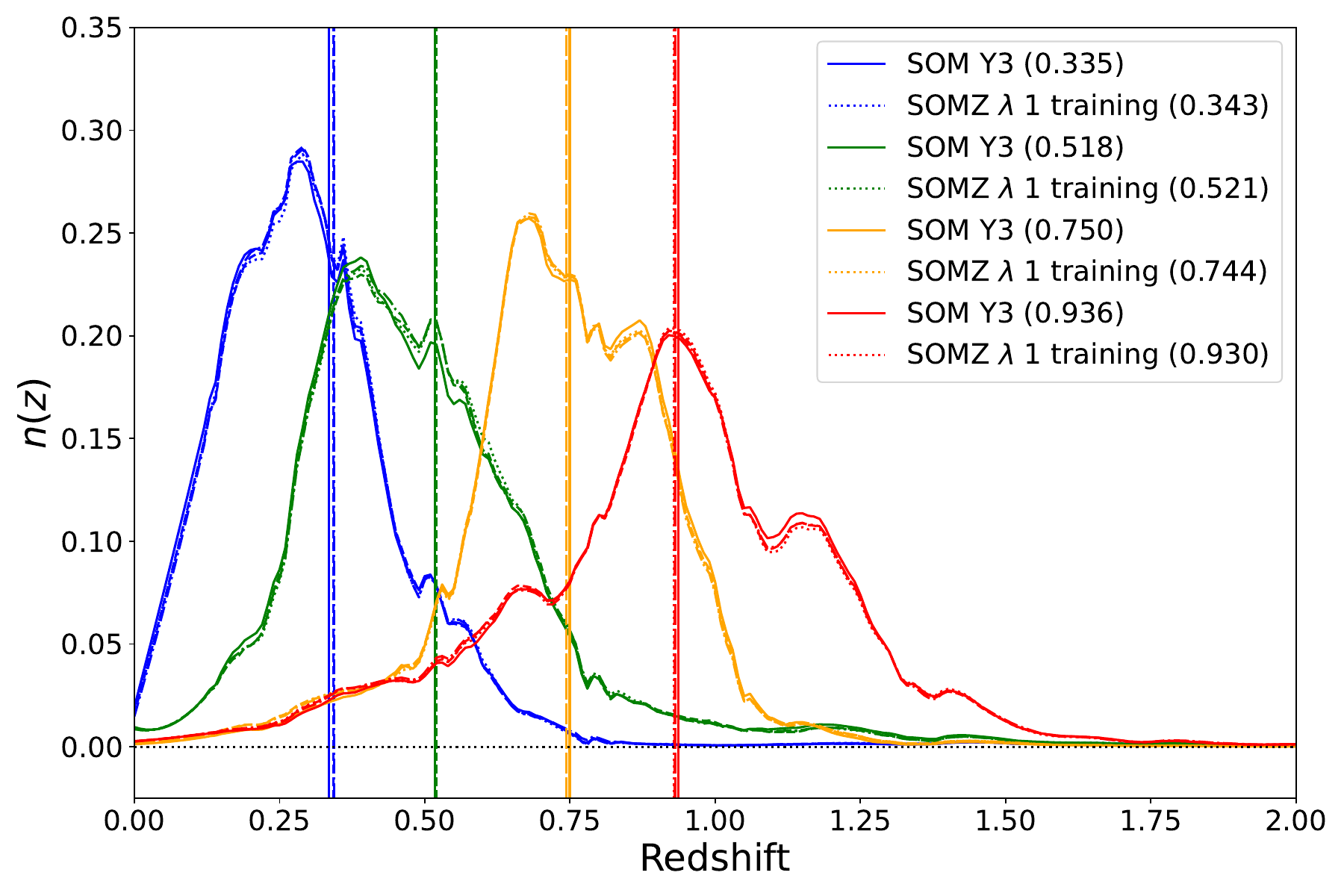}
\caption{Comparison of the \textit{n(z)} bins obtained using the fiducial DES Y3 SOM (solid line), and adding redshift only in the in training phase of the deep SOM. The dotted line represents the most "extreme" case, where $\lambda = 1$ and the contribution of the redshift in training and assigning is the same as the fluxes, while the dashed and dot-dashed lines represent $\lambda = 0.1$ and $\lambda = 0.05$ respectively. The vertical lines are the mean redshift in each bin, shown in the legend for the fiducial method, or $\lambda = 0$,  and the $\lambda=1$ case.}
\label{fig:nz tzaf}
\end{figure}

\begin{figure}
\centering
\includegraphics[width=\linewidth]{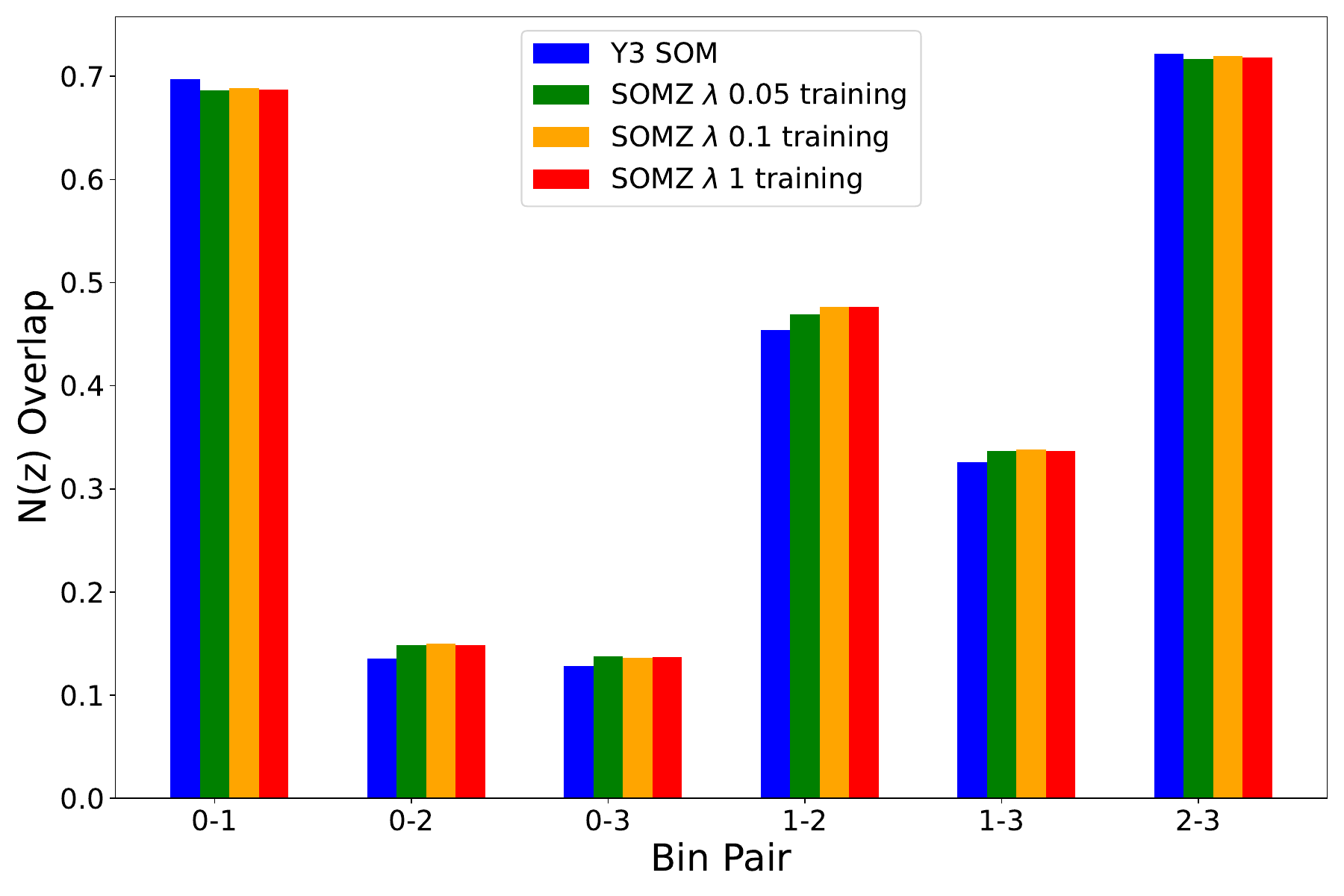}
\caption{Redshfit bin overlap plot for fiducial DES Y3 (blue) and adding redshift in only training the deep SOM. The bin overlap increases as the contribution of the redshift, represented by $\lambda$, increases. The green line represents $\lambda = 0.05$ or $5\%$ contribution, the yellow $\lambda = 0.1$, contributing $10\%$, and the red line $\lambda=1$, contributing the same as flux.}
\label{fig:overlap tzaf}
\end{figure}

Given that adding the redshift information the training and assigning phases of the deep SOM seems to impact negatively our $n(z)$ bins, we conducted a final test before abandoning the concept: adding the redshift information only during the training phase of the deep SOM. That still has a clustering effect upon the deep SOM, as we can see in Figure~\ref{fig:som addz} (bottom), but to a lesser degree. In this case, we observe that the $n(z)$ distribution in each bin, shown in Figure \ref{fig:nz tzaf}, and the bin overlap, shown in Figure \ref{fig:overlap tzaf}, are very similar to those of the fiducial Y3 method, but still slightly worse given that the Y3 SOM still presents the smallest overlap. 

Based on these findings, we infer that incorporating the redshift of individual galaxies as an additional feature alongside fluxes in the estimation of the $n(z)$ distribution using the SOMPZ method is not a viable approach.

\section{Cosmic Shear Measurement}\label{sec:shear_meas}

The small distortions in the observed shapes of galaxies due to weak gravitational lensing by the intervening large scale structure of the Universe are called cosmic shear. Considering two redshift bins $i$ and $j$, the shear correlation function estimator can be written in terms of a galaxy measured tangential, $\epsilon_t$, and radial, $\epsilon_{\times}$, ellipticities as 
\begin{equation}
    \xi_{\pm}^{ij}(\theta) = \langle\epsilon_t\epsilon_t \pm \epsilon_{\times}\epsilon_{\times}\rangle(\theta).
\end{equation}
We can determine the shear-shear statistics by averaging over all galaxy pairs $(a,b)$ separated by and angle $\theta$
\begin{equation}
     \xi_{\pm}^{ij}(\theta) = \frac{\sum_{ab}w_a w_b[\epsilon_t^i\epsilon_t^j \pm \epsilon_{\times}^i\epsilon_{\times}^j]}{\sum_{ab}w_a w_b R_a R_b}
     \label{eq:xipm}
\end{equation}
where $w$ represents the the per-galaxy inverse-variance weight, which is taken over galaxy pairs whose angular separation is within an interval $\Delta\theta$ around $\theta$, and $R$ is the shear response correction from Metacalibration.

The tomographic DES Y3 cosmic shear data vector, $D$, is computed using Equation \ref{eq:xipm} and the code TreeCorr\footnote{\href{https://github.com/rmjarvis/TreeCorr}{https://github.com/rmjarvis/TreeCorr}}. It includes four auto-correlations and six cross-correlations between redshift bins for both positive and negative angular scales, spanning $2.5$ to $250.0$ arcmin. The impact of baryonic effects is mitigated by excluding small angular scales, leaving a total of $167$ $(60)$ data points for $ \xi_{+}$ $( \xi_{-})$ correlations.

The covariance matrix, $C$, is a function of the redshift distributions, cosmological parameters and nuisance parameters. We assume a multivariate Gaussian distribution to model the statistical uncertainties in our cosmic shear data vector. The complete modelling of the disconnected 4-point function part of the covariance matrix is described in \cite*{Friedrich2021}. We compute the connected 4-point function part of the covariance matrix and the contribution from super-sample covariance using the public code CosmoCov\footnote{\href{https://github.com/CosmoLike/CosmoCov}{https://github.com/CosmoLike/CosmoCov.}} 
\cite*{cosmoCov}, which is part of the CosmoLike framework \cite*{cosmolike}.

Following \cite*{DES:2021bvc}, \cite*{DES:2021vln} and previous cosmic shear analysis, we use a iteratively fixed covariance matrix. This means that we start with a set of fiducial input parameters, in our case we use the DES Y3 best fit parameters. Then the covariance is recomputed at the best fit from this first iteration, and the final chains are run.This update had negligible effects on the cosmic shear constraints that we present in this paper.

\section{Modelling and Analysis Choices}\label{sec:theory}

We carry out our analysis in the context of the flat $\Lambda$CDM cosmological model. The cosmological parameters are
$\left\{ \Omega_{\rm m}, \Omega_{\rm b}, h_0, A_{\rm s}, n_{\rm s}, \Omega_{\rm \nu}h^2 \right\}$,
where $\Omega_{\rm m}$ is the density parameter for matter, and $\Omega_{\rm b}$ the equivalent for baryons; $h_0$ is the dimensionless Hubble constant; $A_{\rm s}$ and $n_{\rm s}$ are the amplitude and slope of the primordial curvature power spectrum at a scale of $k=0.05$ Mpc$^{-1}$ respectively; and $\Omega_{\rm \nu}h^2$ is the neutrino mass density parameter. We assume three degenerate massive neutrino species, following \citet*{krause21}.  


\subsubsection{Modelling Cosmic Shear}\label{sec:cosmicshear}

For two redshift bins, $i$ and $j$, the two-point cosmic shear correlations $\xi_\pm^{ij}(\theta)$ can be obtained by decomposing the convergence power spectrum $C_{\kappa}(\ell)$, at an angular wavenumber $\ell$, into $E$- and $B$-mode components \citep*{CrittendenEB, Schneider_vanW_Mell_2002} 

\begin{align} 
    \xi_{+}^{ij}(\theta) &=  \sum_{\ell}\frac{2\ell+1}{4\pi}G_{\ell}^{\pm}\left(\cos\theta\right)\left[C_{\kappa,EE}^{ij}(\ell)+C_{\kappa,BB}^{ij}(\ell)\right], \label{eq: xip Hankel transform}\\ 
    \xi_{-}^{ij}(\theta) &= \sum_{\ell}\frac{2\ell+1}{4\pi}G_{\ell}^{\pm}\left(\cos\theta\right)\left[C_{\kappa,EE}^{ij}(\ell)-C_{\kappa,BB}^{ij}(\ell)\right], \label{eq: xim Hankel transform}
\end{align}
\noindent
where the functions $G^\pm_\ell(x)$ are calculated from Legendre polynomials $P_\ell(x)$ and averaged over angular bins (see Eqs.~19 and 20 in \citealt*{krause21}).

The 2D convergence power spectrum $C^{ij}_\kappa(\ell)$ can be written in terms of the 3D matter power spectrum, assuming the Limber approximation \citep*{Limber53, Limber_LoVerde2008}, as: 
\begin{equation}\label{eq: Limber}
C_{\kappa}^{ij}(\ell)=\int_{0}^{\chi(z_\textrm{max})} \mathrm{d}\chi\frac{W^{i}(\chi)W^{j}(\chi)}{\chi^{2}}P_\delta \left(\frac{\ell+0.5}{\chi},z(\chi)\right),
\end{equation}
\noindent
where $P_\delta(k,z)$  is the nonlinear matter power spectrum and the lensing weight is:
\begin{equation}\label{eq: lensing kernel W}
W^{i}(\chi)=\frac{3H_{0}^{2}\Omega_{\mathrm{m}}}{2c^{2}}\frac{\chi}{a(\chi)}\int_{\chi}^{\chi_{\mathrm{H}}}\mathrm{d}\chi'\,n^{i}\left(z(\chi')\right)\frac{\mathrm{d}z}{\mathrm{d}\chi'}\frac{\chi'-\chi}{\chi'},
\end{equation}
\noindent
with the source galaxy redshift distribution $n^i(z)$ normalised to integrate to 1, and $\chi_\text{H}$ the horizon distance. 
The effect of intrinsic alignments is modelled using the tidal alignment and tidal torquing model (TATT; \citealt*{Blazek_2015}, \citealt*{Blazek_2019})
We follow \citet*{krause21}, and model $P_{\delta}$ using a combination of \blockfont{CAMB} \citep*{lewis00} for the linear part, and \blockfont{HaloFit} \citep*{takahashi12} for nonlinear modifications.
As highlighted in \cite*{DES:2021bvc} and \cite*{DES:2021vln}, the impact of higher order contributions to the observed two-point statistics is verified to be negligible for the scales covered in this work.

\subsubsection{Nuisance Parameters \& Scale Cuts}

Our setup matches the fiducial choices of the DES Y3 cosmic shear analysis. The only significant difference is that, for the sake of simplicity, we choose not to use the additional shear ratio likelihood included by \citet*{DES:2021bvc};  \citet*{DES:2021vln} (a similar decision was made for validating the analysis choices pre-unblinding; see \citealt*{krause21}). As a result, our model space is slightly smaller, since we do not need to vary parameters for galaxy bias or lens photo$-z$ error. The corresponding parameters and their priors are shown in Table \ref{table: priors}. Note that these are identical to the priors used in the Y3 analysis.
We also adopt the fiducial DES Y3 cosmic shear scale cuts (see \citealt*{krause21} for an explanation of how these were derived). 

\begin{table}
	\centering
	\vspace{1cm}
    \caption{A summary of the central values and priors used in our analysis. The top seven rows are cosmological parameters, while those in the lower sections are nuisance parameters corresponding to astrophysics and data calibration. Priors are either uniform (U) or normally-distributed, $\mathcal{N}(\mu,\sigma)$.  
    } \label{table: priors}

\begin{tabular}{c c c }
\hline 
\hline
\bf{Parameter} & \bf{Fiducial Value} & \bf{Prior}\tabularnewline
\hline 
\hline 
\multicolumn{3}{c}{{\bf Cosmological  Parameters}} \\
\omegam & $0.29$ & $\mathrm{U}[0.1, 0.9]$ \tabularnewline
\omegab & $0.052$ & $\mathrm{U} [0.03, 0.07]$ \tabularnewline
$h$     & $0.75$ & $\mathrm{U}[0.55, 0.91]$ \tabularnewline
\as     & $2.38 \times 10^{-9}$ & $\mathrm{U}[0.5,  5.0]\times 10^{-9}$ \tabularnewline
\ns     & $0.99$ & $\mathrm{U}[0.87, 1.07]$   \tabularnewline
$\Omega_{\nu}h^2$ & $0.00053$ & $\mathrm{U}[0.6, 6.44]\times 10^{-3}$   \tabularnewline
\hline
\multicolumn{3}{c}{{\bf Calibration  Parameters}}  \\
$m_{1}$ & $0.0$ & $\mathcal{N}(0.0, 0.0059)$ \tabularnewline
$m_{2}$ & $0.0$ & $\mathcal{N}(0.0, 0.0042)$ \tabularnewline
$m_{3}$ & $0.0$ & $\mathcal{N}(0.0,  0.0054)$ \tabularnewline
$m_{4}$ & $0.0$ & $\mathcal{N}(0.0, 0.0072)$ \tabularnewline
$\Delta z_{1}$ & $0.0$ & $\mathcal{N}(0.0,0.018)$ \tabularnewline
$\Delta z_{2}$ & $0.0$ & $\mathcal{N}(0.0,0.015)$ \tabularnewline
$\Delta z_{3}$ & $0.0$ & $\mathcal{N}(0.0,0.011)$ \tabularnewline
$\Delta z_{4}$ & $0.0$ & $\mathcal{N}(0.0, 0.017)$ \tabularnewline
\hline
\multicolumn{3}{c}{{\bf Intrinsic Alignment Parameters}} \\
$A_1$ & $0.7$ & $\mathrm{U}[-5, 5]$ \tabularnewline
$A_2$ & $-1.36$ & $\mathrm{U}[-5, 5]$ \tabularnewline
$\eta_1$ & $-1.7$ & $\mathrm{U}[-5, 5]$ \tabularnewline
$\eta_2$ & $-2.5$ & $\mathrm{U}[-5, 5]$ \tabularnewline
$b_{\rm TA}$ & $1.0$ & $\mathrm{U}[0, 2]$ \tabularnewline
\hline 
\hline 
\end{tabular}
\end{table}

\subsubsection{Generating Mock Data}\label{sec:data:gen_data}

In this section, we outline the process of generating mock data, which serves as a means to assess the impact of including the \textit{g}-band in the DES Y3 setup. For a set of input parameters, we generate four noiseless DES Y3-like cosmic shear data vectors denoted as $\boldsymbol{D}$. These data vectors are produced using the theoretical pipeline described in Section~\ref{sec:theory} and are centered around the central values outlined in Table \ref{table: priors}.

All four data vectors share the same input flat \lcdm~cosmological model, with parameters set as follows: $\omegam=0.29$, $\as=2.38\times10^{-9}$, $\omegab=0.052$, $h=0.75$, $\ns=0.99$, and $\Omega_{\nu}h^{2}=0.00053$. This configuration corresponds to $\sigma_8=0.79$ and $S_8=0.77$, where $S_8 \equiv \sigma_8\sqrt{\omegam/0.3}$. However, each data vector is distinct in terms of the redshift distribution of the source galaxies, determined using one of the methods employed in this study:
1) One data vector utilizes the redshift distribution obtained by the Y3 SOM, using the \textit{riz} bands.
2) Another data vector adopts the redshift distribution obtained by the SOMF method, utilizing the \textit{riz} bands.
3) A third data vector relies on the redshift distribution derived from the Y3 SOM, employing the \textit{griz} bands.
4) The final data vector is constructed with the redshift distribution acquired through the SOMF method, employing the \textit{griz} bands.
Our analysis framework and mock data generation follow the choices made in the DES Y3, ensuring that our assessments are consistent with the established DES Year 3 standards.

\subsubsection{Bayesian Inference}

For the purpose of parameter estimation, the \textit{likelihood function} of the data vector, $D$, given the model, $T$, characterized by parameters, $\mathbf{p}$, can be represented as $\mathcal{L}(D|\mathbf{p})$. This probability distribution is presumed to follow a multivariate Gaussian distribution
\begin{align}
    \ln \mathcal{L}(D|\mathbf{p}) = - \frac{1}{2}\sum_{i j } \biggr(D_i - T_i(\mathbf{p})\biggr) [C]^{-1}_{ij} \biggr(D_j - T_j(\mathbf{p})\biggr)
\end{align}
$D_i$ represents the $i$th component within the data vector $\xi_{\pm}$, together with its covariance matrix, $C$ (see Section \ref{sec:shear_meas}). Initially, this vector incorporates 20 angular data points, each spanning across the intersections of 4 redshift bins and 2 correlation functions, leading to a total of 227 data points after constraining the angular scales. The corresponding theoretical predictions for these statistical quantities, represented as $T_i(\mathbf{p})$, are elaborated upon in this section. The Bayesian \textit{posterior probability} distributions of the cosmological parameters, denoted as $\mathcal{P}(\mathbf{p}|D)$, are derived by combining the likelihood with the \textit{prior probabilities}, $P(\mathbf{p})$, as outlined in Table \ref{table: priors}, following the principles of Bayes’ theorem
\begin{equation}
    \mathcal{P}(\mathbf{p}|D) = \frac{P(\mathbf{p})\mathcal{L}(D|\mathbf{p})}{P(D)}
\end{equation}
where $P(D)$ is the \textit{evidence} of the data.

The posterior distribution is sampled using the Polychord \citep*{Handley:2015a, Handley:2015b}. The analysis framework is based on CosmoSIS \citep*{zuntz15}, a modular tool for estimating cosmological parameters. We use the fiducial sampler settings (500 live points, tolerance 0.01) that have been verified to showcase the precision of the posterior distributions and Bayesian evidence estimations (as discussed in \citealt*{y3-samplers}).

\section*{Affiliations}


 $^{1}$ Department of Physics, Carnegie Mellon University, Pittsburgh, Pennsylvania 15312,  USA \\
 $^{2}$ NSF AI Planning Institute for Physics of the Future, Carnegie Mellon University, Pittsburgh, PA 15213,  USA \\
 $^{3}$ Department of Physics, Duke University Durham, NC 27708,  USA \\
 $^{4}$ Institute of Astronomy, University of Cambridge, Madingley Road, Cambridge CB3 0HA, UK \\
 $^{5}$ Kavli Institute for Cosmology, University of Cambridge, Madingley Road, Cambridge CB3 0HA, UK \\
 $^{6}$ Argonne National Laboratory, 9700 South Cass Avenue, Lemont, IL 60439,  USA \\
 $^{7}$ Institute of Space Sciences (ICE, CSIC),  Campus UAB, Carrer de Can Magrans, s/n,  08193 Barcelona, Spain \\
 $^{8}$ Department of Physics and Astronomy, University of Pennsylvania, Philadelphia, PA 19104,  USA \\
 $^{9}$ Kavli Institute for Cosmological Physics, University of Chicago, Chicago, IL 60637,  USA \\
 $^{10}$ Institut de F\'{\i}sica d'Altes Energies (IFAE), The Barcelona Institute of Science and Technology, Campus UAB, 08193 Bellaterra (Barcelona) Spain \\
 $^{11}$ Department of Astrophysical Sciences, Princeton University, Peyton Hall, Princeton, NJ 08544,  USA \\
 $^{12}$ Department of Physics, Northeastern University, Boston, MA 02115,  USA \\
 $^{13}$ Department of Physics, University of Michigan, Ann Arbor, MI 48109,  USA \\
 $^{14}$ Physics Department, 2320 Chamberlin Hall, University of Wisconsin-Madison, 1150 University Avenue Madison, WI  53706-1390 \\
 $^{15}$ Laborat\'orio Interinstitucional de e-Astronomia - LIneA, Rua Gal. Jos\'e Cristino 77, Rio de Janeiro, RJ - 20921-400, Brazil \\
 $^{16}$ Instituto de F\'{i}sica Te\'orica, Universidade Estadual Paulista, S\~ao Paulo, Brazil \\
 $^{17}$ Brookhaven National Laboratory, Bldg 510, Upton, NY 11973,  USA \\
 $^{18}$ Universidad de La Laguna, Dpto. Astrofísica, E-38206 La Laguna, Tenerife, Spain \\
 $^{19}$ Instituto de Astrofisica de Canarias, E-38205 La Laguna, Tenerife, Spain \\
 $^{20}$ Department of Astronomy, University of Illinois at Urbana-Champaign, 1002 W. Green Street, Urbana, IL 61801,  USA \\
 $^{21}$ Center for Astrophysical Surveys, National Center for Supercomputing Applications, 1205 West Clark St., Urbana, IL 61801,  USA \\
 $^{22}$ Physics Department, William Jewell College, Liberty, MO, 64068 \\
 $^{23}$ Department of Astronomy and Astrophysics, University of Chicago, Chicago, IL 60637,  USA \\
 $^{24}$ NASA Goddard Space Flight Center, 8800 Greenbelt Rd, Greenbelt, MD 20771,  USA \\
 $^{25}$ Jodrell Bank Center for Astrophysics, School of Physics and Astronomy, University of Manchester, Oxford Road, Manchester, M13 9PL, UK \\
 $^{26}$ Kavli Institute for Particle Astrophysics \& Cosmology, P. O. Box 2450, Stanford University, Stanford, CA 94305,  USA \\
 $^{27}$ Lawrence Berkeley National Laboratory, 1 Cyclotron Road, Berkeley, CA 94720,  USA \\
 $^{28}$ Fermi National Accelerator Laboratory, P. O. Box 500, Batavia, IL 60510,  USA \\
 $^{29}$ Universit\'e Grenoble Alpes, CNRS, LPSC-IN2P3, 38000 Grenoble, France \\
 $^{30}$ Jet Propulsion Laboratory, California Institute of Technology, 4800 Oak Grove Dr., Pasadena, CA 91109,  USA \\
 $^{31}$ Department of Astronomy/Steward Observatory, University of Arizona, 933 North Cherry Avenue, Tucson, AZ 85721-0065,  USA \\
 $^{32}$ Department of Physics and Astronomy, University of Waterloo, 200 University Ave W, Waterloo, ON N2L 3G1, Canada \\
 $^{33}$ Department of Astronomy, University of California, Berkeley,  501 Campbell Hall, Berkeley, CA 94720,  USA \\
 $^{34}$ SLAC National Accelerator Laboratory, Menlo Park, CA 94025,  USA \\
 $^{35}$ University Observatory, Faculty of Physics, Ludwig-Maximilians-Universit\"at, Scheinerstr. 1, 81679 Munich, Germany \\
 $^{36}$ School of Physics and Astronomy, Cardiff University, CF24 3AA, UK \\
 $^{37}$ Department of Astronomy, University of Geneva, ch. d'\'Ecogia 16, CH-1290 Versoix, Switzerland \\
 $^{38}$ Department of Physics, University of Arizona, Tucson, AZ 85721,  USA \\
 $^{39}$ Department of Applied Mathematics and Theoretical Physics, University of Cambridge, Cambridge CB3 0WA, UK \\
 $^{40}$ Instituto de F\'isica Gleb Wataghin, Universidade Estadual de Campinas, 13083-859, Campinas, SP, Brazil \\
 $^{41}$ Nordita, KTH Royal Institute of Technology and Stockholm University, Hannes Alfv\'ens v\"ag 12, SE-10691 Stockholm, Sweden \\
 $^{42}$ Department of Physics, University of Genova and INFN, Via Dodecaneso 33, 16146, Genova, Italy \\
 $^{43}$ ICTP South American Institute for Fundamental Research\\ Instituto de F\'{\i}sica Te\'orica, Universidade Estadual Paulista, S\~ao Paulo, Brazil \\
 $^{44}$ Center for Cosmology and Astro-Particle Physics, The Ohio State University, Columbus, OH 43210,  USA \\
 $^{45}$ Space Telescope Science Institute, 3700 San Martin Drive, Baltimore, MD  21218,  USA \\
 $^{46}$ Centro de Investigaciones Energ\'eticas, Medioambientales y Tecnol\'ogicas (CIEMAT), Madrid, Spain \\
 $^{47}$ Department of Physics and Astronomy, Stony Brook University, Stony Brook, NY 11794,  USA \\
 $^{48}$ Institut de Recherche en Astrophysique et Plan\'etologie (IRAP), Universit\'e de Toulouse, CNRS, UPS, CNES, 14 Av. Edouard Belin, 31400 Toulouse, France \\
 $^{49}$ Excellence Cluster Origins, Boltzmannstr.\ 2, 85748 Garching, Germany \\
 $^{50}$ Max Planck Institute for Extraterrestrial Physics, Giessenbachstrasse, 85748 Garching, Germany \\
 $^{51}$ Universit\"ats-Sternwarte, Fakult\"at f\"ur Physik, Ludwig-Maximilians Universit\"at M\"unchen, Scheinerstr. 1, 81679 M\"unchen, Germany \\
 $^{52}$ Department of Physics, Stanford University, 382 Via Pueblo Mall, Stanford, CA 94305,  USA \\
 $^{53}$ Cerro Tololo Inter-American Observatory, NSF's National Optical-Infrared Astronomy Research Laboratory, Casilla 603, La Serena, Chile \\
 $^{54}$ Institute for Astronomy, University of Edinburgh, Edinburgh EH9 3HJ, UK \\
 $^{55}$ Institute of Cosmology and Gravitation, University of Portsmouth, Portsmouth, PO1 3FX, UK \\
 $^{56}$ Department of Physics \& Astronomy, University College London, Gower Street, London, WC1E 6BT, UK \\
 $^{57}$ Institut d'Estudis Espacials de Catalunya (IEEC), 08034 Barcelona, Spain \\
 $^{58}$ Institute for Fundamental Physics of the Universe, Via Beirut 2, 34014 Trieste, Italy \\
 $^{59}$ Astronomy Unit, Department of Physics, University of Trieste, via Tiepolo 11, I-34131 Trieste, Italy \\
 $^{60}$ INAF-Osservatorio Astronomico di Trieste, via G. B. Tiepolo 11, I-34143 Trieste, Italy \\
 $^{61}$ Institute of Theoretical Astrophysics, University of Oslo. P.O. Box 1029 Blindern, NO-0315 Oslo, Norway \\
 $^{62}$ Instituto de Fisica Teorica UAM/CSIC, Universidad Autonoma de Madrid, 28049 Madrid, Spain \\
 $^{63}$ School of Mathematics and Physics, University of Queensland,  Brisbane, QLD 4072, Australia \\
 $^{64}$ Santa Cruz Institute for Particle Physics, Santa Cruz, CA 95064, USA \\
 $^{65}$ Department of Physics, The Ohio State University, Columbus, OH 43210, USA \\
 $^{66}$ Center for Astrophysics $\vert$ Harvard \& Smithsonian, 60 Garden Street, Cambridge, MA 02138, USA \\
 $^{67}$ Australian Astronomical Optics, Macquarie University, North Ryde, NSW 2113, Australia \\
 $^{68}$ Lowell Observatory, 1400 Mars Hill Rd, Flagstaff, AZ 86001, USA \\
 $^{69}$ Departamento de F\'isica Matem\'atica, Instituto de F\'isica, Universidade de S\~ao Paulo, CP 66318, S\~ao Paulo, SP, 05314-970, Brazil \\
 $^{70}$ George P. and Cynthia Woods Mitchell Institute for Fundamental Physics and Astronomy, and Department of Physics and Astronomy, Texas A\&M University, College Station, TX 77843,  USA \\
 $^{71}$ LPSC Grenoble - 53, Avenue des Martyrs 38026 Grenoble, France \\
 $^{72}$ Instituci\'o Catalana de Recerca i Estudis Avan\c{c}ats, E-08010 Barcelona, Spain \\
 $^{73}$ Observat\'orio Nacional, Rua Gal. Jos\'e Cristino 77, Rio de Janeiro, RJ - 20921-400, Brazil \\
 $^{74}$ Hamburger Sternwarte, Universit\"{a}t Hamburg, Gojenbergsweg 112, 21029 Hamburg, Germany \\
 $^{75}$ Ruhr University Bochum, Faculty of Physics and Astronomy, Astronomical Institute, German Centre for Cosmological Lensing, 44780 Bochum, Germany \\
 $^{76}$ Physics Department, Lancaster University, Lancaster, LA1 4YB, UK \\
 $^{77}$ Computer Science and Mathematics Division, Oak Ridge National Laboratory, Oak Ridge, TN 37831 \\
 $^{78}$ Argonne National Laboratory, 9700 S Cass Ave, Lemont, IL 60439, USA \\

\end{document}